\documentclass[longauth]{aa}  

\usepackage{graphicx}
\usepackage{xcolor}
\usepackage{tikz}
\usepackage{txfonts}
\newcommand{\kms}{\rm km\,s^{-1}}
\newcommand{\Msun}{\rm M_\odot}
\newcommand{\re}{r_{\rm e}}
\newcommand{\rt}{r_{\rm t}}
\newcommand{\va}{v_{\rm a}}
\newcommand{\vc}{v_{\rm circ}}
\newcommand{\vdisp}{\sigma_0}
\newcommand{\Mdyn}{M_{\rm dyn}}
\newcommand{\Mgas}{M_{\rm gas}}
\newcommand{\Mbar}{M_{\rm bar}}
\newcommand{\micron}{\rm \mu m}
\definecolor{purple}{RGB}{76, 0,153}

\begin{document}

   \title{Ionised gas kinematics and dynamical masses of $z\gtrsim6$ galaxies from JADES/NIRSpec high-resolution spectroscopy
   }

   \author{Anna de Graaff\inst{\ref{i1},\thanks{degraaff@mpia.de}}
          \and  Hans-Walter Rix\inst{\ref{i1}}
          \and Stefano Carniani\inst{\ref{i3}}
          \and Katherine A. Suess\inst{\ref{i14},\ref{i18}}
          \and St\'ephane Charlot\inst{\ref{i5}}
          \and Emma Curtis-Lake\inst{\ref{i2}}
          \and Santiago Arribas \inst{\ref{i10}}
          \and William M. Baker\inst{\ref{i6},\ref{i7}} 
          \and Kristan Boyett\inst{\ref{i20},\ref{i21}}
          \and Andrew J.\ Bunker\inst{\ref{i4}}
          \and Alex J. Cameron\inst{\ref{i4}}
          \and Jacopo Chevallard\inst{\ref{i4}}
          \and Mirko Curti\inst{\ref{i23},\ref{i6},\ref{i7}}
          \and Daniel J.\ Eisenstein\inst{\ref{i13}}
          \and Marijn Franx\inst{\ref{i19}}
          \and Kevin Hainline\inst{\ref{i16}}
          \and Ryan Hausen\inst{\ref{i22}}
          \and Zhiyuan Ji\inst{\ref{i16}} 
          \and Benjamin D.\ Johnson\inst{\ref{i13}}
          \and Gareth C. Jones\inst{\ref{i4}}
          \and Roberto Maiolino\inst{\ref{i6},\ref{i7},\ref{i8}}
          \and Michael V. Maseda\inst{\ref{i11}}
          \and Erica Nelson\inst{\ref{i17}} 
          \and Eleonora Parlanti\inst{\ref{i3}} 
          \and Tim Rawle\inst{\ref{i12}}
          \and Brant Robertson\inst{\ref{i14}}
          \and Sandro Tacchella\inst{\ref{i6},\ref{i7}}
          \and Hannah \"Ubler\inst{\ref{i6},\ref{i7}}
          \and Christina C. Williams\inst{\ref{i15}}
          \and Christopher N. A. Willmer\inst{\ref{i16}} 
          \and Chris Willott\inst{\ref{i9}}
          }

   \institute{Max-Planck-Institut f\"ur Astronomie, K\"onigstuhl 17, 69117 Heidelberg, Germany\label{i1}
    \and
    Scuola Normale Superiore, Piazza dei Cavalieri 7, I-56126 Pisa, Italy\label{i3}  
    \and              
    Department of Astronomy and Astrophysics University of California, Santa Cruz, 1156 High Street, Santa Cruz CA 96054, USA\label{i14}   
    \and
    Kavli Institute for Particle Astrophysics and Cosmology and Department of Physics, Stanford University, Stanford, CA 94305, USA\label{i18} 
    \and
    Sorbonne Universit\'e, CNRS, UMR 7095, Institut d'Astrophysique de Paris, 98 bis bd Arago, 75014 Paris, France\label{i5}  
    \and    
    Centre for Astrophysics Research, Department of Physics, Astronomy and Mathematics, University of Hertfordshire, Hatfield AL10 9AB, UK\label{i2}   
    \and
    Centro de Astrobiolog\'ia (CAB), CSIC–INTA, Cra. de Ajalvir Km.~4, 28850- Torrej\'on de Ardoz, Madrid, Spain\label{i10}  
    \and    
    Kavli Institute for Cosmology, University of Cambridge, Madingley Road, Cambridge, CB3 0HA, UK\label{i6}   
    \and
    Cavendish Laboratory - Astrophysics Group, University of Cambridge, 19 JJ Thomson Avenue, Cambridge, CB3 0HE, UK\label{i7}   
    \and
    School of Physics, University of Melbourne, Parkville 3010, VIC, Australia\label{i20}  
    \and
    ARC Centre of Excellence for All Sky Astrophysics in 3 Dimensions (ASTRO 3D), Australia\label{i21}  
    \and
    Department of Physics, University of Oxford, Denys Wilkinson Building, Keble Road, Oxford OX1 3RH, UK\label{i4}  
    \and
    European Southern Observatory, Karl-Schwarzschild-Strasse 2, 85748 Garching, Germany\label{i23}
    \and
    Center for Astrophysics $|$ Harvard \& Smithsonian, 60 Garden St., Cambridge MA 02138 USA\label{i13}  
    \and    
    Leiden Observatory, Leiden University, P.O.Box 9513, NL-2300 AA Leiden, The Netherlands\label{i19}
    \and
    Steward Observatory, University of Arizona, 933 N. Cherry Avenue, Tucson, AZ 85721, USA\label{i16}  
    \and
    Department of Physics and Astronomy, The Johns Hopkins University, 3400 N. Charles St., Baltimore, MD 21218\label{i22}
    \and
    Department of Physics and Astronomy, University College London, Gower Street, London WC1E 6BT, UK\label{i8}  
    \and
    Department of Astronomy, University of Wisconsin-Madison, 475 N. Charter St., Madison, WI 53706 USA\label{i11}  
    \and
    Department for Astrophysical and Planetary Science, University of Colorado, Boulder, CO 80309, USA\label{i17} 
    \and
    European Space Agency (ESA), European Space Astronomy Centre (ESAC), Camino Bajo del Castillo s/n, 28692 Villafranca del Castillo, Madrid, Spain\label{i12}  
    \and
    NSF’s National Optical-Infrared Astronomy Research Laboratory, 950 North Cherry Avenue, Tucson, AZ 85719, USA\label{i15}  
    \and
    NRC Herzberg, 5071 West Saanich Rd, Victoria, BC V9E 2E7, Canada\label{i9}  
             }


\titlerunning{Resolving galaxy kinematics at $z>6$ with JWST/NIRSpec}
\authorrunning{de Graaff et al.}


  \abstract{
  We explore the kinematic gas properties of six $5.5<z<7.4$ galaxies in the JWST Advanced Deep Extragalactic Survey (JADES), using high-resolution JWST/NIRSpec multi-object spectroscopy of the rest-frame optical emission lines [O{\sc iii}] and H$\alpha$. The objects are small and of low stellar mass ($\sim 1\,$kpc; $M_*\sim10^{7-9}\,\Msun$), less massive than any galaxy studied kinematically at $z>1$ thus far. The cold gas masses implied by the observed star formation rates are $\sim 10\times$ larger than the stellar masses. We find that their ionised gas is spatially resolved by JWST,  with evidence for broadened lines and spatial velocity gradients. Using a simple thin-disc model, we fit these data with a novel forward modelling software that accounts for the complex geometry, point spread function, and pixellation of the NIRSpec instrument. We find the sample to include both rotation- and dispersion-dominated structures, as we detect velocity gradients of $v(\re)\approx100-150\,\kms$, and find velocity dispersions of $\vdisp\approx 30-70\,\kms$ that are comparable to those at cosmic noon. The dynamical masses implied by these models ($\Mdyn\sim10^{9-10}\,\Msun$) are larger than the stellar masses by up to a factor 40, and larger than the total baryonic mass (gas + stars) by a factor of $\sim 3$. Qualitatively, this result is robust even if the observed velocity gradients reflect ongoing mergers rather than rotating discs.
  Unless the observed emission line kinematics is dominated by outflows, this implies that the centres of these galaxies are dark-matter dominated or that star formation is $3\times$ less efficient, leading to higher inferred gas masses. 
  }

   \keywords{Galaxies: evolution -- Galaxies: high-redshift -- Galaxies: kinematics and dynamics -- Galaxies: structure}

   \maketitle
%
\section{Introduction}

In the nearby Universe galaxies show a variety of dynamical structures and structural components, which are reflective of their mass assembly histories \citep[e.g.][]{Cappellari2016,vdSande2018,Falcon2019}. However, the details of the formation and evolution of these structures -- nominally, rotationally supported discs and spheroidal bulges supported primarily by dispersion -- are still unclear. The physical conditions in the early Universe, the secular evolution of galaxies, and mergers with other systems are all likely to play important roles. One outstanding question in particular is when and how early galaxies settled into dynamically cold discs. 

Although this question is ideally answered by measuring spatially resolved stellar kinematics across cosmic time, such measurements have only been possible up to $z\sim1$ \citep[e.g,][]{vanHoudt2021}, except for a few strongly lensed, massive galaxies at $z\sim2$ \citep{Newman2018}. Instead, the ionised gas of the interstellar medium (ISM) provides critical insight into the dynamical properties of (star-forming) galaxies across a much wider redshift range \citep[for a review, see][]{FSchreiber2020}. Many studies have focused on inferring galaxy dynamical properties from rest-frame optical emission lines, to map the evolution in the velocity dispersion ($\sigma$) and the ratio between the rotation velocity and dispersion ($v/\sigma$), which measures the degree of rotational support of the system. 

Measurements of the ionised gas kinematics from multiple large spectroscopic surveys of star-forming galaxies at $z\sim 1-4$ have demonstrated that the velocity dispersion of star-forming galaxies increases with redshift, while the rotational support decreases to $v/\sigma\approx1$ by $z\approx 3$ \citep[e.g.,][]{Wisnioski2015,Wisnioski2019,Stott2016,Simons2017,Turner2017,Price2020}. In addition, imaging studies have shown that galaxy morphologies are less disc-like and more clumpy at rest-frame UV wavelengths at higher redshifts \citep[e.g.,][]{vdWel2014,Guo2015,Zhang2019,Sattari2023}. Theoretical models and simulations have suggested that gravitational instabilities, the accretion of gas and smaller systems from the cosmic web, and stellar feedback may be responsible for increased turbulence at higher redshifts \citep[e.g.,][]{Dekel2009b,Genel2012,Ceverino2012,Krumholz2018}. 

However, using submillimeter observations from the Atacama Large Millimeter Array (ALMA), several studies have found dynamically cold discs at $z\sim2-6$ \citep[e.g.,][]{Neeleman2020,Jones2021,Lelli2021,Rizzo2021,Parlanti2023,Pope2023}, even finding $v/\sigma \approx 20$ at $z\approx 4.5$ \citep{Fraternali2021}. These observations raise the question of how such systems formed and settled so rapidly (within $\sim1$\,Gyr), and how these observations can be reconciled with the aforementioned studies at cosmic noon. However, the ALMA observations infer the galaxy kinematics from far-infrared and millimetre transitions (CO, [C{\sc ii}]), which trace colder gas than the rest-frame optical lines, likely explaining part of the discrepancy \citep{Uebler2019,Rizzo2023}. Additionally, selection effects likely play an important role, as many of the ALMA observations primarily probe the most massive galaxies at $z>4$.

To understand the evolution of more typical ($\sim M^*$) galaxies, requires spatially resolved spectroscopy of faint galaxies at high redshifts. In this regime, ground-based telescopes are unable to observe rest-frame optical emission lines, whereas sub-mm facilities are in principle able to observe such systems, but at extremely high cost. The launch of the James Webb Space Telescope (JWST) has enabled spectroscopy with very high sensitivity and high spatial resolution \citep{Gardner2023,Rigby2023}. Using the slitless spectroscopy mode of JWST/NIRCam \citep{NIRCamInst}, Nelson et al. ({in prep.}) reveal the ionised gas kinematics in a massive galaxy at $z\approx5$. However, only the NIRSpec instrument provides the spectral resolution needed to resolve galaxy kinematics for low-mass systems \citep[$\sigma\approx 50\,\kms$ for a uniformly-illuminated slit;][]{Jakobsen2022}. JWST/NIRSpec additionally provides a multi-object spectroscopic (MOS) mode \citep{Ferruit2022}, allowing for the simultaneous observation of up to $\approx 200$ objects, making observations of high-redshift targets highly efficient. The slit-based observations with the microshutter array (MSA) however sacrifice one spatial dimension with respect to integral field spectroscopy \citep[IFS;][see also \citealt{Price2016}]{Boeker2022}. Therefore, extra care is required to extract spatial and dynamical information from NIRSpec MSA data.

In this paper, we present the dynamical properties of six high-redshift galaxies ($z>5.5$) in the JWST Advanced Deep Extragalactic Survey \citep[JADES;][]{Eisenstein2023}. These objects are spatially extended in deep NIRCam imaging and were followed up with the high-resolution NIRSpec MOS mode, providing spatially-resolved spectroscopy of their rest-frame optical emission lines. The data are presented in Section~\ref{sec:data}. To model the galaxy kinematics, we propagate analytical models through a simulated NIRSpec instrument, and use MCMC sampling to fit the data (Section~\ref{sec:method}), using NIRCam imaging as a prior on the morphology. We present the results of our modelling in Section~\ref{sec:results}, demonstrating a diverse range of kinematic structures in a previously unexplored population of galaxies. We discuss the possibility that some systems may be late-stage mergers in Section~\ref{sec:discussion}, and examine the large discrepancy between the derived dynamical masses and stellar masses. We summarise our findings in Section~\ref{sec:conclusion}.

\section{Data}\label{sec:data}

\subsection{NIRSpec spectroscopy}

We use NIRSpec MOS observations in the GOODS-South field taken as part of the JADES deep and medium programmes \citep[ID numbers 1210 and 1286, PI N. L\"utzgendorf;][]{Bunker2023,Eisenstein2023}. Targets were selected from a combination of JWST/NIRCam and Hubble Space Telescope (HST) imaging and followed up with JWST/NIRSpec using the low-resolution prism ($R\sim100$), the three medium-resolution gratings ($R\sim1000$), and the reddest high-resolution grating (G395H; $R\sim2700$ for a uniformly illuminated slit). Here, we focus primarily on the high-resolution spectroscopy, although we also use the prism data to estimate stellar masses and star formation rates (SFRs). The spectra in our sample were obtained using a 3-point nodding pattern and vary in depth, ranging from 2.2\,h to 7.0\,h of total integration time for the G395H grating (summarised in Table~\ref{tab:texp}). The total exposure times for the prism range from 2.2\,h to 28\,h.

The NIRSpec data were reduced using the NIRSpec GTO collaboration pipeline (Carniani et al. in prep), as is also described in \citet{CurtisLake2023} and \citet{Bunker2023}. Crucially, in contrast with other studies so far using NIRSpec data, our analysis largely does not rely on the final rectified, combined, and extracted 2D and 1D spectra generated by the pipeline. Instead, we perform our dynamical modelling to intermediate data products: we use 2D cutouts of the detector from individual exposures that have been background-subtracted and flat-fielded. In this way, we mitigate correlated noise and artificial broadening that is otherwise introduced by the rectification and combination algorithm of the reduction pipeline. We also note that these intermediate data products do not correct for any slit losses due to the spatial extent of the sources, as these effects are already fully accounted for in our modelling (Section~\ref{sec:method}).

Nevertheless, the 2D rectified high-resolution spectra and their 1D extractions were used for the initial visual inspection and selection of the sample. We also used the 1D extracted prism data for spectral energy distribution (SED) modelling in Section~\ref{sec:results} to estimate stellar masses and SFRs.

\subsection{NIRCam imaging}\label{sec:nircam}
s
Even though some objects were initially selected based on HST imaging, JWST/NIRCam imaging is available for all targets in our sample from a combination of Cycle 1 programmes. The majority of our targets fall within the JADES footprint in GOODS-S \citep{Rieke2023}, and are therefore imaged in 9 different NIRCam filters. 
One of the selected targets (JADES-NS-10016374) is located outside of the JADES footprint but falls within the footprint of the FRESCO survey \citep[programme 1895, PI P. Oesch;][]{FRESCO2023}. Although FRESCO is primarily a grism survey, the survey also obtained imaging with three different NIRCam filters (F182M, F210M, F444W), although at significantly reduced exposure times compared with JADES imaging. 

All images were reduced as described in \citet{Rieke2023}. We use the JWST Calibration Pipeline v1.9.6 with the CRDS pipeline mapping (pmap) context 1084. We run Stage 1 and Stage 2 of the pipeline with the default parameters, but provided our own sky-flat for the flat-fielding. Following Stage 2, we perform a custom subtraction of the 1/f noise, scattered-light effects (“wisps”) and the large-scale background. We perform an astrometric alignment using a custom version of JWST TweakReg, aligning our images to the HST F814W and F160W mosaics in the GOODS-S field with astrometry tied to Gaia-EDR3 (G. Brammer priv. comm.). We achieve an overall good alignment with relative offsets between bands of less than 0.1 short-wavelength pixel (< 3 mas). We then run Stage 3 of the JWST pipeline, combining all exposures of a given filter and a given visit.

\begin{table*}[!h]
    \caption{ Coordinates and G395H exposure times of the selected sample. }
    \centering
    \begin{tabular}{l l l l l l l}
    \hline \hline
     JADES ID & NIRSpec ID\tablefootmark & R.A. & Dec. & $N_{\rm exp}$ & $t_{\rm exp, total}$ & line  \\
         & & (deg) &  (deg) & & (ksec) & \\ 
        \hline
       JADES-GS+53.13002-27.77839 & JADES-NS-00016745 & 53.13005	& -27.77839  & 18 & 25.2 & H$\alpha$\\
       JADES-GS+53.17655-27.77111 & JADES-NS-00019606 & 53.17654 &	-27.77111 & 6 & 8.4 & [O{\sc iii}]\\
       JADES-GS+53.15407-27.76607 & JADES-NS-00022251 & 53.15407 & -27.76608 & 12 & 16.8 & [O{\sc iii}] \\
       JADES-GS+53.18343-27.79097 & JADES-NS-00047100\tablefootmark & 53.18343	& -27.79098 & 9 & 8.0 & [O{\sc iii}]\\
       JADES-GS+53.11572-27.77495  & JADES-NS-10016374 & 53.11572 &	-27.77496 & 12 & 16.8 & H$\alpha$\\
       JADES-GS+53.18374-27.79390 & JADES-NS-20086025 &  53.18375 & -27.79389 & 9 & 8.0 & [O{\sc iii}]\\
    \hline
    \end{tabular}
    \tablefoot{
    \tablefoottext{a}{The ID number corresponds to the NIRSpec ID as described in \citet{Bunker2023}.}
    \tablefoottext{b}{Best-fit coordinates from the morphological modelling to the NIRCam imaging. These differ slightly from the coordinates in the JADES ID due to updated astrometry.}
    }
    \label{tab:texp}
\end{table*}

For our analysis we select the NIRCam filter that most closely represents the emission line morphology of the target. Given the high equivalent widths of the emission lines in our sample, we use the medium band covering the emission line where available (four out of six objects). For the remaining two objects we instead use broad-band filters. {We use the available low-resolution prism spectra to quantify the flux originating from emission lines versus the stellar continuum in Appendix~\ref{sec:beagle}. For the four objects with medium-band images, we hence estimate that $\approx 70-75\%$ of the NIRCam flux is due to emission lines, and therefore provide a good map of the ionised gas. For the two objects with broad-band images the stellar continuum dominates (emission line fluxes contribute $\approx 35\%$), and we discuss how this may affect the inferred kinematic parameters in Sections~\ref{sec:results} and \ref{sec:mass_budget}.} 

\subsection{Sample selection}\label{sec:sample}

We have visually inspected all (358) JADES targets in GOODS-S for which high-resolution NIRSpec spectroscopy is available thus far. We select objects that are at high redshift ($z>5.5$, i.e. when the Universe was less than 1\,Gyr old), are spatially extended in the NIRCam imaging ($r_{\rm e}\gtrsim0.1\arcsec$), and have bright emission lines, i.e. an integrated signal-to-noise ratio (SNR) $\gtrsim 20$. Additionally, we require that the 1D spectrum shows no obvious evidence for a broad-line component, which would be indicative of large-scale outflows \citep[i.e., the sample discussed in][]{Carniani2023}. The resulting sample consists of six objects that span the redshift range $z= 5.5 - 7.4$. 
We observe H$\alpha$ in two of these objects, and [O{\sc iii}] in the remaining four. 
For the two highest redshift targets H$\alpha$ is outside the wavelength coverage of NIRSpec ($z>7$); for two $z\sim6$ objects we do not observe H$\alpha$, as the traces of the high-resolution spectra are long and therefore fall partially off the detector.

\begin{figure*}[!h]
    \centering
    \includegraphics[width=0.87\linewidth]{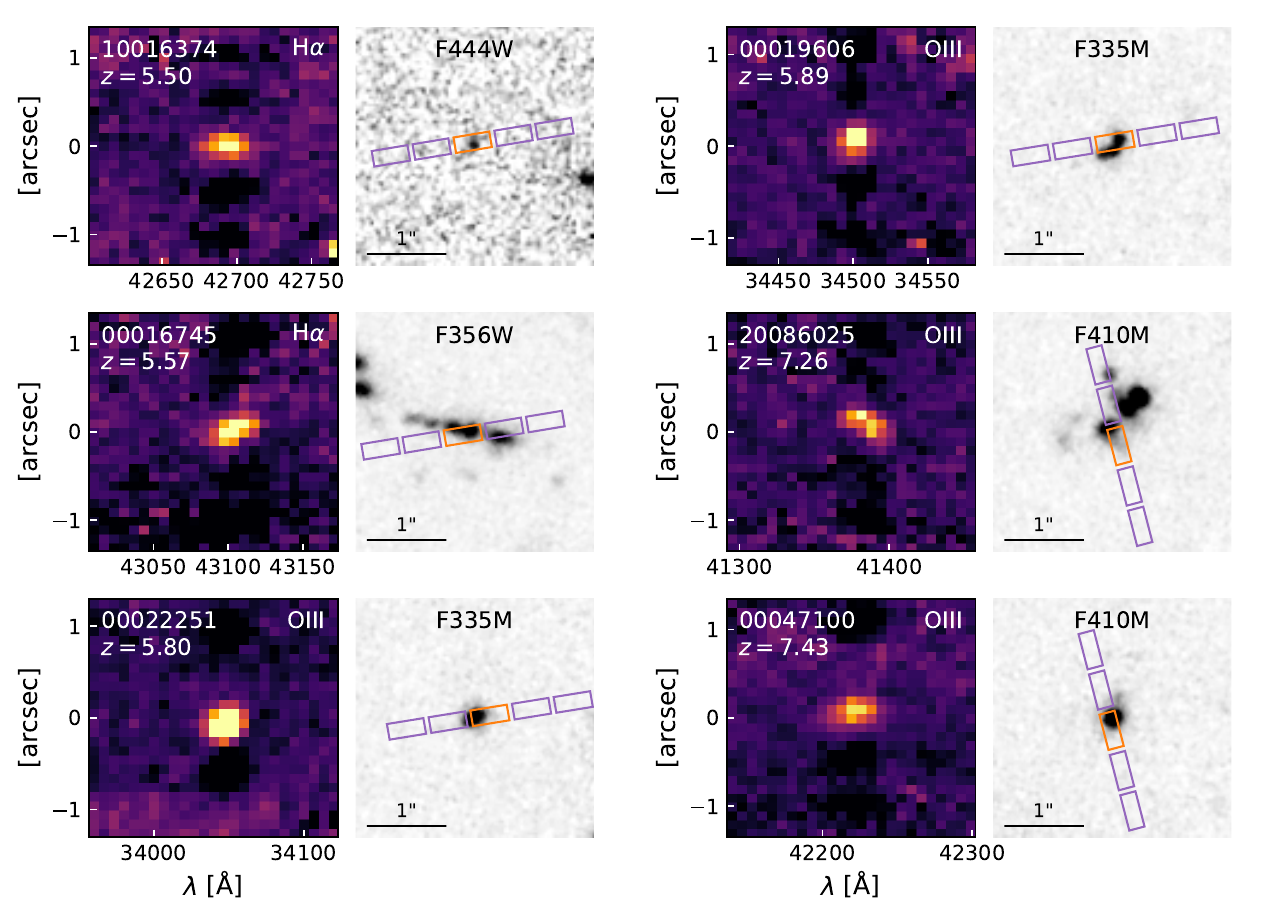}
    \caption{Sample of six spatially-resolved high-redshift objects in JADES. Left panels show cutouts of the emission lines in the 2D rectified and combined spectra obtained with the high-resolution G395H grating. Negatives in the cutouts are the result of the background subtraction method used. Right panels show NIRCam image cutouts for each object (JADES, FRESCO), for the band that most closely resembles the emission line morphology (Section~\ref{sec:nircam}). The 3-shutter slits and 3-point nodding pattern used result in an effective area of 5 shutters: the shutter encompassing the source is shown in orange, and the shutters used for background subtraction are shown in purple.  }
    \label{fig:sample_overview}
\end{figure*}

The 2D combined and rectified images of the emission lines are presented in Fig.~\ref{fig:sample_overview} together with cutouts from the NIRCam imaging, where we have selected the NIRCam filter closest to the emission line as described in the previous section. The positions of the microshutters are shown in orange. 

\begin{figure*}[!h]
    \centering
    \includegraphics[width=0.87\linewidth]{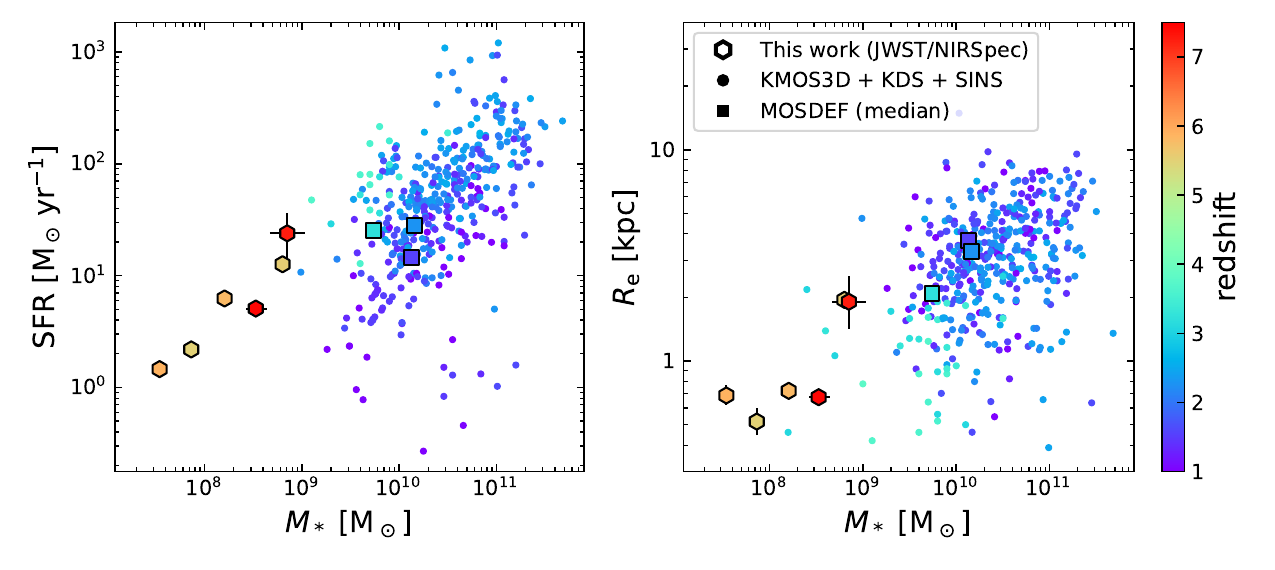}
    \caption{Sample distribution in the parameter space of the stellar mass, SFR and emission line half-light radius, colour-coded by the redshift. We compare with a selection of ground-based near-infrared spectroscopic surveys that used the H$\alpha$ and [O{\sc iii}] lines to measure galaxy kinematics at $z\sim1-4$ \citep{Turner2017,SINS2018,Wisnioski2019,Price2020}; we use UV+IR SFRs from \citet{Whitaker2014} where available for the \citet{Turner2017} sample. We note that all objects at $z>1$ from KMOS3D are shown in this figure, but not all objects were used in the kinematic studies due to low SNR or their size being too small. The sample selected from JADES (hexagons) probes a very different regime: JADES targets are at higher redshift and lower stellar mass than ground-based facilities have probed thus far.  }
    \label{fig:param_space}
\end{figure*}

We note that, given the complex selection function of JADES \citep{Bunker2023}, and our additional imposed selection criteria during the visual inspection, the selected sample is far from complete in (stellar) mass, magnitude or star formation rate. However, our aim is to demonstrate the ability of NIRSpec/MOS to measure galaxy kinematics in an entirely new regime: Fig.~\ref{fig:param_space} shows that the targets in our sample are not only at a higher redshift than has been attainable so far with ground-based near-infrared spectrographs but are also significantly smaller and less massive. %
We defer a comprehensive analysis of the galaxy kinematics as a function of redshift and mass to a future paper, as this will require the complete JADES and NIRSpec WIDE datasets \citep[][Maseda et al. in prep.]{Eisenstein2023} as well as a thorough understanding of the selection functions of the different survey tiers.

\section{Dynamical modelling} \label{sec:method}

We use a Bayesian forward-modelling approach to estimate the dynamical properties of the galaxies from 2D emission spectra, observed through the NIRSpec MSA apertures. First, we construct parametric model cubes for the flux distribution ($I(\vec{x},\lambda)$) based on analytical surface brightness and velocity profiles. Second, we model the complexities of the NIRSpec instrument that are imprinted on the data when mapping the kinematic models onto a mock NIRSpec detector. Third, we use a Markov chain Monte Carlo (MCMC) sampling method to fit the models to the spectra, adopting S\'ersic profile fits to NIRCam imaging as a prior. We defer a detailed description of this forward-modelling and fitting software (\texttt{msafit}) to a future paper, in which we will also demonstrate convergence tests and comparison with calibration data. In this section, we only provide a summary overview of the models and software, which we release publicly together with this paper\footnote{Software, reference data (PSF models, trace models) and installation instructions can be found here: \url{https://github.com/annadeg/jwst-msafit}}.

\subsection{Thin disc models}\label{sec:model_description}

Although the spatial and spectral resolution of JWST/NIRSpec are high, the small systems in our sample are close to the resolution limit. This suggests that we should limit ourselves to a relatively simple geometric and dynamical model: a thin rotating disc. Although geometrically thin, we allow for the disc to be kinematically warm by adding a velocity dispersion profile. We discuss the possible limitations of our model choice in Section~\ref{sec:discussion}.

We model the spatial flux distribution of the emission line as a S\'ersic profile \citep{Sersic1968}, 
which is described by four parameters: the total flux $F$, half-light radius $\re$ (major axis), projected minor-to-major axis ratio $q$, and S\'ersic index $n$. As we assume a thin disc model, the projected axis ratio is directly related to the inclination angle ($i$) of the system. In addition, there are three important position-dependent parameters that enter the model, the position angle (PA) with respect to the MSA slitlet as measured from the positive $x$-axis (i.e. $\rm PA=90\,\deg$ represents perfect alignment with the slit), and the centroid position of the object within the shutter (${\rm d}x$, ${\rm d}y$).

For the velocity field we use the common empirical description of an arctangent rotation curve \citep{Courteau1997}:
\begin{equation}
    v(r) = \frac{2}{\pi}v_{\rm a} \arctan\left({\frac{r}{r_{\rm t}}} \right)\,,
\end{equation}
where $v_{\rm a}$ is the asymptotic or maximum velocity with respect to the systemic velocity of the galaxy, and $r_{\rm t}$ is the turnover radius. We parametrise the systemic velocity as the mean wavelength of the emission line $\lambda_0$. To allow for a kinematically warm disc, we additionally assume a constant velocity dispersion profile $\sigma(r)\equiv\sigma_0$ across the disc.

Combined, this amounts to a 11-parameter model ($\lambda_0$, $F$, ${\rm d}x$, ${\rm d}y$, $\re$, $n$, $q$, PA, $r_{\rm t}$, $\va$, $\vdisp$). We convolve the flux profile and velocity profiles to form a model flux cube $I(x,y,\lambda)$, where $x$ and $y$ are the coordinates in the MSA plane and sampled in user-specified intervals. To construct these cubes, we ensure that the spatial and wavelength dimensions are sampled at minimum at Nyquist frequency for the point spread function (PSF) at the wavelength considered or the NIRSpec pixel size ($0.1\arcsec$), whichever is smaller. However, this sampling would be too sparse to evaluate the steep S\'ersic profiles at small radii. In order to accurately integrate S\'ersic profiles we therefore first oversample the spatial grid dynamically, such that the innermost region ($<0.2\,\re$) is oversampled by a factor 500 and the outer regions ($>\re$) by a factor 10, and then integrate the profile onto the coarser grid.

\subsection{Forward modelling the NIRSpec MSA}\label{sec:msa_description}

\begin{figure*}
    \centering
    \includegraphics[width=\linewidth]{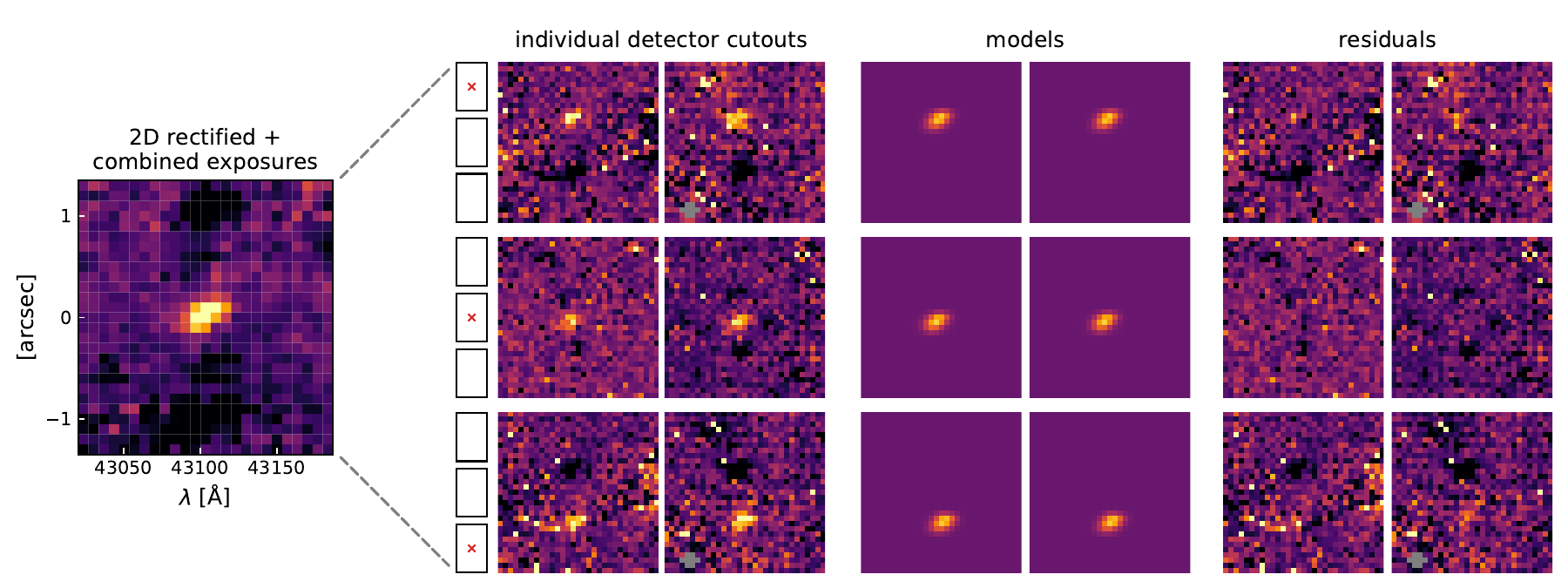}
    \caption{Example of the fitting procedure for object JADES-NS-00016745 (Fig.~\ref{fig:sample_overview}). Although the final combination of all exposures (left) was used for our initial visual inspection and sample selection, the pixels in this spectrum are highly correlated. Instead of using this combined spectrum, we simultaneously fit to all individual exposures obtained. In the case of JADES-NS-00016745 two exposures were taken per nodding position, resulting in six independent measurements for one 3-point nodding pattern with NIRSpec. To combat the undersampled PSF of NIRSpec, we perform our modelling in the detector plane, propagating parametric models to the exact same location on the detector as the observed data. The likelihood is then computed from the combination of all residual images. Pixels flagged by the reduction pipeline as affected by cosmic rays are masked and shown in grey. }
    \label{fig:method}
\end{figure*}

Although forward-modelling software for slit-based multi-object spectroscopy has been developed before \citep{Price2016}, there are several unique challenges to modelling NIRSpec MOS data: (i) the diffraction-limited PSF of JWST, which enables high spatial resolution, but is highly complex in shape; (ii) the complex geometry of the NIRSpec MSA, comprising $\approx 2\times 10^5$ microshutters that are separated by shutter walls, imprints additional diffraction patterns ($+$-shaped due to the slit aperture) as well as shadows on the detector (``bar shadows''); (iii) the relatively large pixels ($\approx 0.1\arcsec\times 0.1\arcsec$) of the NIRSpec detector imply that the PSF is undersampled at all wavelengths\footnote{For a comprehensive technical description of NIRSpec, we refer the reader to \citet{Jakobsen2022} and \citet{Ferruit2022}}.

As a consequence of the first and second above challenges, there is a strong variation of the PSF shape within the shutter. High-redshift extragalactic objects are often also comparable in size to the NIRSpec PSF width and pixel size, and the position of the flux centroid within the shutter therefore strongly affects the shape of the flux distribution on the detector. Moreover, the shutter walls ($\approx0.035\arcsec$) are relatively small compared to the pixel size, making the effects of the bar shadows complex to model. Lastly, the open area of the microshutters is relatively small \citep[$\approx 0.2\arcsec \times 0.46\arcsec$;][]{Jakobsen2022} compared to the PSF size. Slit losses are therefore substantial, even at the centre of the shutter.

We attempt to capture all these effects within our modelling. First, we construct libraries of synthetic PSF models at a range of different wavelengths and spatial offsets, with the PSF centres sampling the shutter every $0.02\arcsec$ and using a 5x oversampling factor for the PSF images. These PSFs represent the 2D image of a point source with an infinitely narrow emission line in the detector plane, and hence contains both the spatial distribution along the slit and the distribution in the wavelength direction. We constructed the PSFs using custom Fourier optical simulations, tracing monochromatic point sources through NIRSpec to the detector focal plane. These models capture the combined PSF of JWST and NIRSpec, including the diffraction and  light losses (often referred to as path losses) caused by the masking by the micro-shutter slitlets and spectrograph pupil. We defer a detailed description to de Graaff et al. (in prep.), but note that the construction of these PSFs is largely the same as in previous works that presented or used the NIRSpec Instrument Performance Simulator \citep{Piqueras2008,Piqueras2010,Giardino2019,Jakobsen2022}. The main difference is that the implementation used for this work is python-based and makes use of the Physical Optics Propagation in python \citep[POPPY;][]{Perrin2012} libraries, allowing a carefully tuned wavelength-independent sampling in both the image and pupil planes. Although these models are based on in-flight calibrations where possible, a number of necessary reference files were created pre-launch. {We therefore caution that there is likely to be a systematic uncertainty in the true width and shape of these PSFs. Unfortunately, neither sufficient nor dedicated calibration data currently exist. We discuss the current status of calibrations in more detail in Appendix~\ref{sec:lsf}, and estimate a $\approx10-20\%$ systematic uncertainty in the PSF full width at half maxima (FWHM; both spatial and spectral) of our models. }

Second, we construct libraries of spectral traces for all microshutters and dispersers, using the instrument model of \citet{Dorner2016} and \citet{Giardino2016}, the parameters of which were tuned during the in-flight commissioning phase \citep{Luetzgendorf2022,Alves2022}. These traces provide a mapping from the centre of a given shutter ($s_{ij}$) and chosen wavelength to the detector plane ($X$, $Y$). We also use this model to derive the tilt angle of the slitlets with respect to the trace in the detector plane  (a few degrees for G395H).

Third, we construct model detectors, for which the pixels are initially oversampled by a factor 5. To reduce computational cost we do not model the full $2048\times2048$ detectors, but create cutouts of $\approx 30\times 30$ pixels around a region of interest, whilst keeping track of the corresponding detector coordinates.

With these libraries and models in place, we can forward model the analytical flux cube of Section~\ref{sec:model_description}. As the PSF strongly varies with the intrashutter position, the model cube cannot be convolved with a single PSF. Instead, we treat the model cube as a collection of point sources, hence propagating each point in the cube with its local PSF. The slices of the model cube are then projected onto the oversampled model detector using the trace library, given a defined shutter ($s_{ij}$). Finally, the detector is downsampled by a factor 5 to match the true detector pixel size, and convolved with a $3\times3$ kernel to mimic the effects of inter-pixel capacitive coupling\footnote{The IPC kernels are publicly available on the JDox: \url{https://jwst-docs.stsci.edu/jwst-near-infrared-spectrograph/nirspec-instrumentation/nirspec-detectors/nirspec-detector-performance}}. This results in a noiseless model image of the input data cube on the detector.

\subsection{Model fitting}\label{sec:fitting_description}

\begin{figure*}
    \centering
    \begin{tikzpicture}[      
            every node/.style={anchor=south west,inner sep=0pt},
            x=1cm, y=1cm,
          ]   
         \node (fig1) at (0,0)
           {\includegraphics[width=0.95\linewidth]{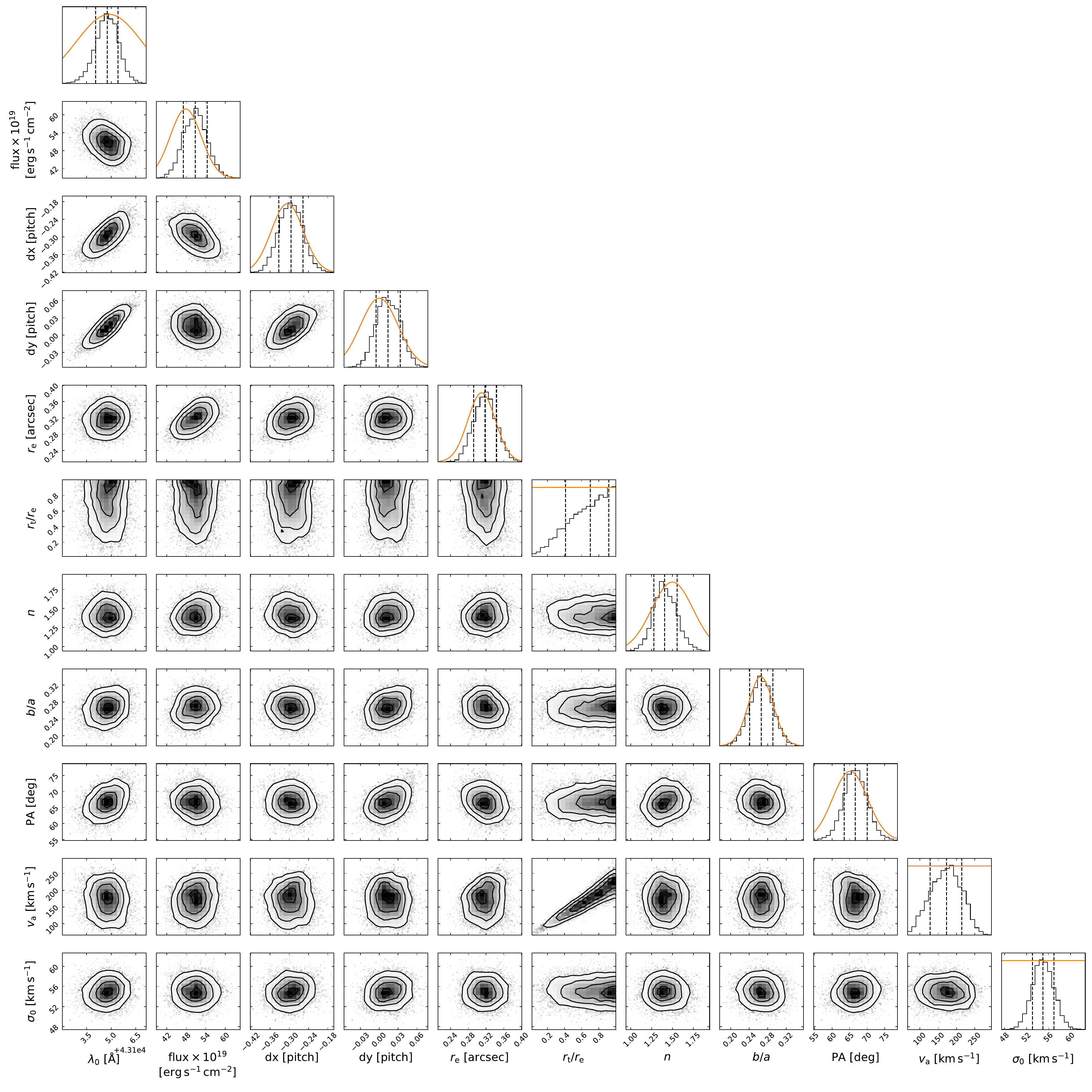}};
         \node (fig2) at (10.95,10.95)
           {\includegraphics[width=0.35\linewidth]{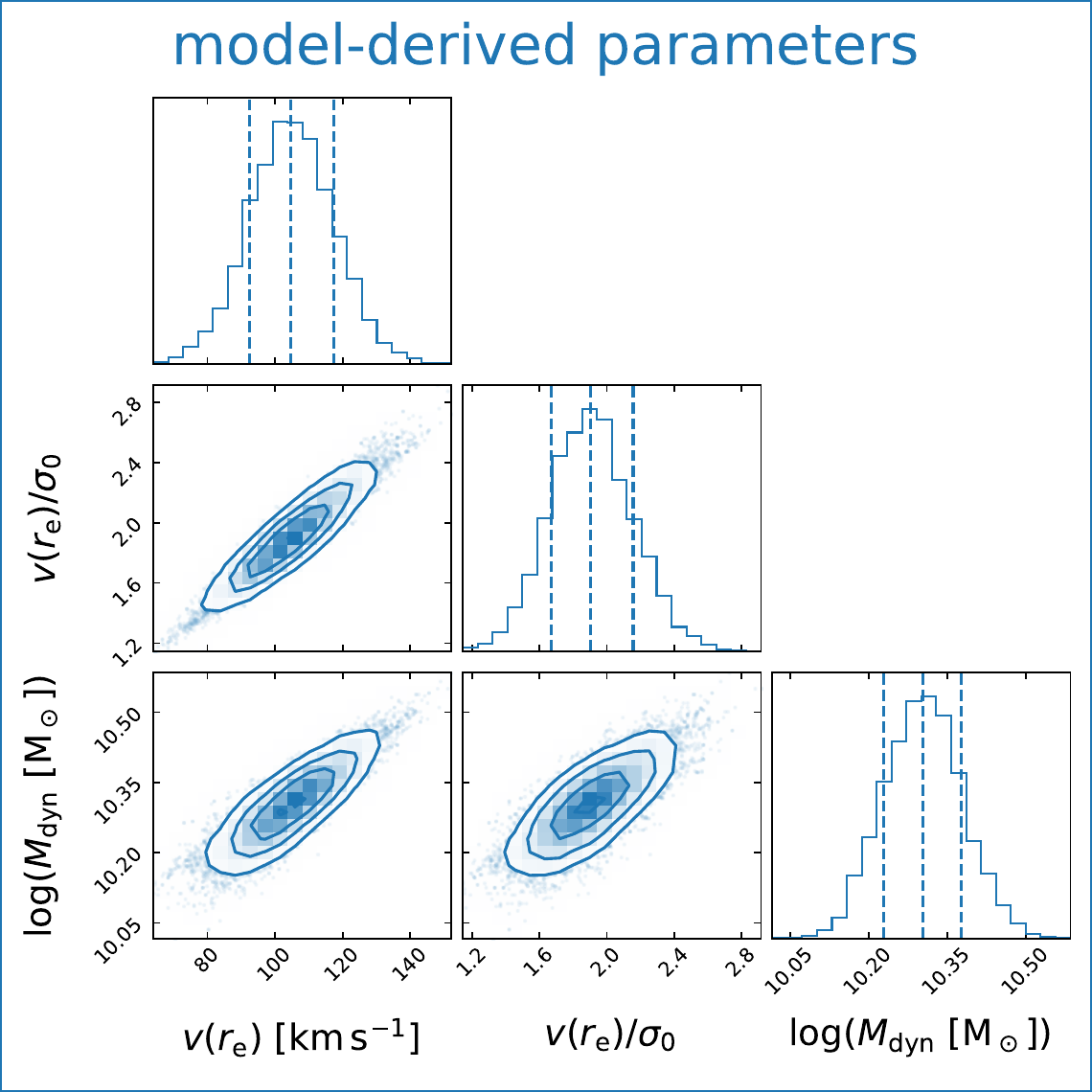}};  
    \end{tikzpicture}

    \caption{Corner plot for the 11-parameter thin-disc model for the object in Fig.~\ref{fig:method}. Histograms show the posterior probability distributions, with orange lines indicating the prior probability distributions. We generally find good constraints on the kinematic parameters $\va$ and $\vdisp$, although the turnover radius $\rt$ is poorly constrained and degenerate with the rotational velocity. The top right panels show (in blue) the parameters that we derive from the model and are discussed in Section~\ref{sec:results}. }
    \label{fig:corner}
\end{figure*}

The procedure of Section~\ref{sec:msa_description} generates a model for a single set of parameters. To estimate the posterior probability distributions of the parameters, we use the MCMC ensemble sampler implemented in the \texttt{emcee} package \citep{emcee}. 

Importantly, we perform the comparison between the models and data in the detector plane to mitigate correlated noise. We hence do not use the 2D combined spectrum, but simultaneously fit to multiple exposures (Fig.~\ref{fig:method}), while masking pixels that are flagged as being affected by cosmic rays or hot pixels. The likelihood function for a set of parameters $\vec{p_{\rm model}}$ then is
\begin{equation}
    \ln \mathcal{L}(F_j | \vec{p_{\rm model}} ) = \sum_i^N -0.5 \left( \sum_j^K \frac{(F_j - M_j(\vec{p_{\rm model}})^2}{\sigma_j} + \ln \left( 2 \pi \sigma_j^2 \right) \right )\,,
\end{equation}
where $N$ is the number of exposures, $K$ the number of unmasked pixels per exposure, and $F_j$, $M_j$ and $\sigma_j$ are the observed flux, model flux, and uncertainty in the $j^{\rm th}$ pixel, respectively.

In calculating the posterior, we use informative priors where possible, as the geometry is poorly constrained based on the spectroscopic data alone. We perform morphological modelling to the NIRCam images (Fig.~\ref{fig:sample_overview}) with \texttt{lenstronomy} \citep{Birrer2018,Birrer2021}, following the procedure described in \citet{Suess2023}. Based on the SNR of the object in the image and corresponding typical uncertainties in the structural parameters derived by \citet{vdWel2012}, we set Gaussian priors centred on the best-fit \texttt{lenstronomy} estimates. To allow for uncertainty in the PSF models and deviations between the image morphology and emission line morphology {(as described in Section~\ref{sec:nircam}; Appendix~\ref{sec:beagle})}, we double all uncertainties in the structural parameters to set the dispersions of the Gaussian priors. The mean wavelength and integrated flux of the emission line can be determined from the 1D spectrum with high accuracy. This line flux needs a correction for the slit losses, which we estimate based on the best-fit \texttt{lenstronomy} model parameters. We then create Gaussian priors for the central wavelength and flux, somewhat conservatively assuming an uncertainty of half a pixel ($\approx 2\AA$) for the wavelength, and a $10\%$ uncertainty in the total flux on the detector. Lastly, we allow for a small uncertainty on the intrashutter position of the source due to the finite pointing accuracy of JWST, for which we use Gaussian priors with a dispersion of $25\,$mas, which is the typical pointing accuracy after the MSA target acquisition \citep{Boeker2023}.

For the parameters of the dynamical model, the maximum velocity ($\va$) and the velocity dispersion ($\vdisp$), we use uniform priors. For the fitting we parametrise the turnover radius as the ratio $r_{\rm t}/\re$, and assume a uniform prior for this ratio. We show an example fit in Figs.~\ref{fig:method} and \ref{fig:corner}, which demonstrates that $\va$ and $\vdisp$ are formally well-constrained ($>5\,\sigma$ significance). The turnover radius is poorly constrained, and forms the largest source of uncertainty on $\va$ due to the degeneracy between $\va$ and $\rt/\re$. This is likely due to a combination of the moderate spatial resolution and the limited spatial extent probed by the microshutters. In Section~\ref{sec:results} we instead compute the rotational velocity at $\re$, $v(\re)$, which is better constrained by the data (blue distributions in the top right of Fig.~\ref{fig:corner}).

We note that the relatively large spatial extent of the source compared to the shutter size may also lead to a loss of flux, as the background subtraction step in the reduction pipeline subtracts flux from the (neighbouring) source that falls in the adjacent shutter. To test the magnitude of this potential bias, we also perform our modelling on a separate reduction that excludes exposures in which the source falls in the central shutter and includes only the outer nods, therefore mitigating any self-subtraction and contamination (but at the cost of a slightly lower overall SNR). We find that the recovered flux is consistent within the error bars with the fit to the standard reduction that uses all 3 nodding positions. Our model is therefore robust against slight self-subtraction present in the spectra, likely helped by the prior information provided by the NIRCam imaging and the relatively small spatial extent of the sources compared to the shutter size.

\section{Results} \label{sec:results}

\subsection{Ionised gas kinematics at $z>5.5$}\label{sec:results_vsigma}

\begin{figure}
    \centering
    \includegraphics[width=\linewidth]{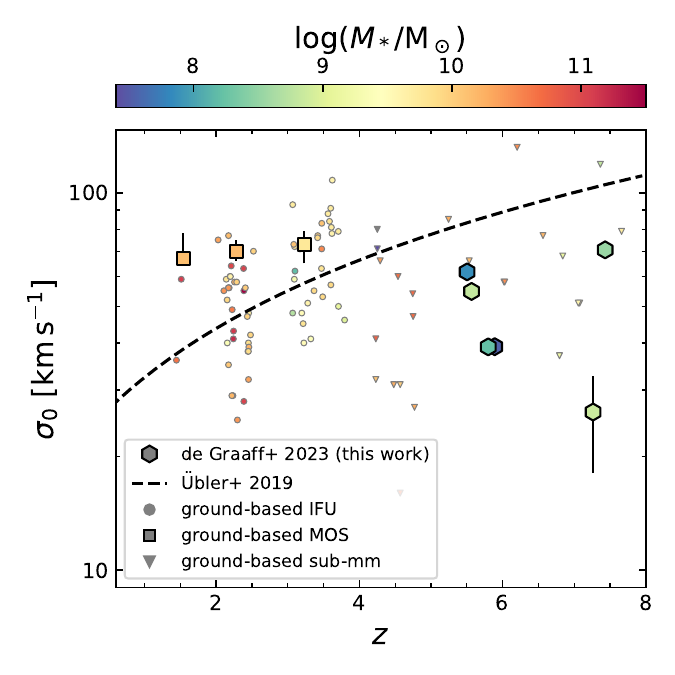}
    \caption{Velocity dispersion of the ionised gas as a function of redshift. The dashed line shows the fit from \citet{Uebler2019} for ionised gas at $0.6<z<2.6$ extrapolated to higher redshifts, while circles show results from a selection of ground-based IFU surveys in the near-infrared \citep{Turner2017,SINS2018} and squares the results from ground-based near-infrared MOS data \citep[the resolved and aligned sample of][]{Price2020}. Blue triangles show results from various studies with ALMA, which measure the kinematics of the cold gas for massive, dusty star-forming galaxies \citep{Neeleman2020,Rizzo2020,Fraternali2021,Lelli2021,Rizzo2021,Herrera2022,Parlanti2023}. The high-redshift objects observed with JWST are dynamically approximately equally turbulent to the observations of more massive galaxies at cosmic noon, and do not follow the trend of increasing $\vdisp$ with redshift observed at $z<4$.}
    \label{fig:vdisp}
\end{figure}

\begin{figure}
    \centering
    \includegraphics[width=\linewidth]{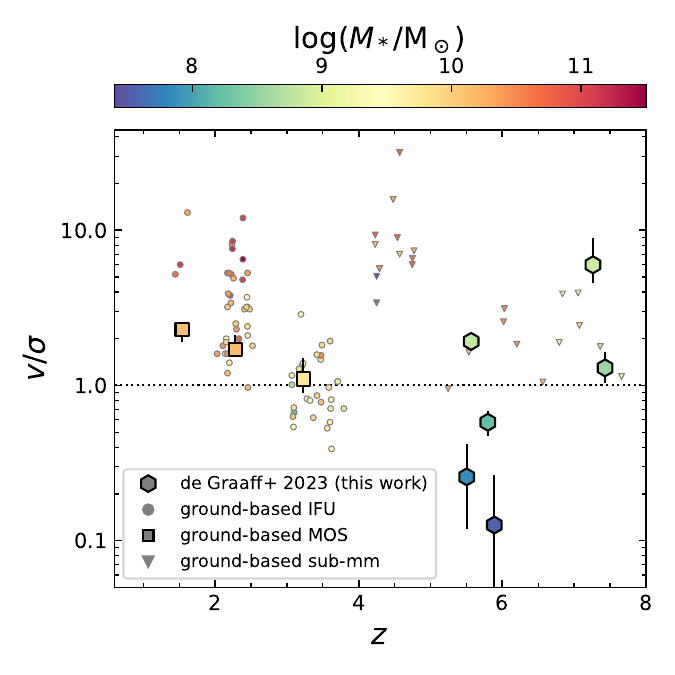}
    \caption{Rotational support as a function of redshift, measured as the ratio between the velocity at the effective radius and the constant velocity dispersion: $v(\re)/\vdisp$. Although studies based on ground-based near-infrared data (as described in Fig.~\ref{fig:vdisp}) have found a clear, gradual decline in $v/\sigma$ toward higher redshift, we find an interesting diversity among our sample of low-mass galaxies, with dynamically-cold discs existing possibly as early as $z\approx7$. }
    \label{fig:vsigma}
\end{figure}

We present the results of our modelling for the sample of six objects in Tables~\ref{tab:fit_results} and \ref{tab:fit_results2}. Significant rotation is detected in three of the objects, and marginally detected or consistent with zero in the other cases. As we did not remove galaxies that are strongly misaligned with the slit, some of these objects (e.g., object JADES-NS-00019606) may have a velocity gradient that is simply not observable at this position angle of the MSA. Nevertheless, the measurement of the velocity dispersion is still useful in these cases, and also provides confirmation that our model is able to return $\va\approx0$ despite our a priori assumption that the system is rotating. 

We find five objects have velocity dispersions that are broader than the instrument line-spread function (LSF) for a point source ($\sigma_{\rm inst}\approx 25-30\,\kms$, see Appendix~\ref{sec:lsf}), which is the relevant LSF after accounting for the source morphology. The formal uncertainties on the dispersions are small, which is likely due to our assumption of a thin disc, meaning that our estimate of the error on $\sigma_0$ does not include an uncertainty caused by the degeneracy between the true disc thickness and inclination angle. The sixth object (JADES-NS-20086025) has a highly complex morphology with apparent extended diffuse (line) emission. Due to the relatively low SNR of this emission and the crowded region around the target, the morphological fit to the F410M image with \texttt{lenstronomy} for this object does not converge. To model the dynamics we therefore significantly weaken the morphological priors for this object: we set a uniform prior on the S\'ersic index, and set only weak constraints on the effective radius (Gaussian prior of $r_{\rm e}=0.3\pm0.3\arcsec$), axis ratio (Gaussian prior of $q=0.3\pm0.2$) and position angle ($\rm PA=90\pm15\deg$). We do find convergence for the dynamical model, although several of the inferred parameters have larger uncertainties than the other objects in our sample.

Fig.~\ref{fig:vdisp} shows the velocity dispersions of our sample as a function of redshift. We compare with a selection of ground-based near-infrared studies: the MOS results of \citet[][the medians of their resolved and aligned sample]{Price2020}, the IFU samples of \citet{Turner2017} and \citet{SINS2018}, and the analytical fit to the large IFU survey KMOS3D at $z\sim1-3$ from \citet{Uebler2019}. 

Extrapolating the fit by \citet{Uebler2019} at $z\sim1-3$ to $z\sim 7$ suggests that the ionised gas in galaxies is highly turbulent at early epochs. In contrast, we find that all objects lie well below this extrapolation, and instead have velocity dispersions that are approximately equal to the average dispersion at $z\sim2-3$. On the other hand, the typical stellar mass of our sample is substantially lower than the literature data (Fig.~\ref{fig:param_space}). If the velocity dispersion depends on stellar mass \citep[as predicted by simulations, e.g.][]{Pillepich2019}, the ISM in these low-mass systems may still have a relatively high turbulence.

We also compare with studies that used ALMA to resolve galaxy kinematics at the same redshift as our sample \citep[blue triangles;][]{Neeleman2020,Rizzo2020,Fraternali2021,Lelli2021,Rizzo2021,Herrera2022,Parlanti2023}. Although these objects lie at the same redshift, the measurements differ substantially: the galaxies observed with ALMA are often more massive ($M_*\gtrsim10^{10}\,\Msun$), and the observed emission lines tend to trace much colder gas. Interestingly, despite these differences, the ALMA-based velocity dispersions are very similar to our measurements of the ionised gas based on rest-frame optical emission lines. Possibly, this is because the effects of the higher mass and lower gas temperature on the velocity dispersion act in opposite directions. Observations of the same systems with both ALMA and JWST will be crucial to constrain these effects.

Next, we compute the ratio $v/\sigma \equiv v(\re)/\vdisp$ and examine its dependence on redshift in Fig.~\ref{fig:vsigma}, comparing with the same literature as mentioned previously. Studies around cosmic noon showed a clear, gradual decline in the degree of rotational support toward higher redshift.
Based on these measurements, one may expect none of the $z>5$ galaxies to be rotation-dominated ($v/\sigma >1$). Yet, we find an interesting diversity among our sample, with three objects having $v/\sigma >1$ even at the highest redshifts ($z\approx7$). We discuss in Section~\ref{sec:discussion} whether these objects may truly form cold rotating discs, or whether these reflect velocity gradients within systems that are not virialised.

We again compare our sample with the ALMA-based studies, which are all rotation-dominated systems with relatively high $v/\sigma$ ratios. Our sample shows greater diversity, which may be due to the fact that the gas tracers differ and the mass range probed is significantly different. The misalignment of some objects with the microshutters also may underestimate the intrinsic $v/\sigma$ ratio of some systems. Larger samples are therefore required to fully understand the different kinematic properties of the gas phases traced with ALMA and JWST at high redshifts 

{Lastly, we revisit Section~\ref{sec:nircam}, where we described that for two out six objects the NIRCam image used as a prior in the emission line modelling predominantly traces stellar continuum emission instead of line emission. If the morphology of the emission line differs strongly from the continuum, this may bias the inferred kinematic parameters, especially if the galaxy is prolate instead of our assumed oblate thin disc model. One of these two objects (JADES-NS-00016745; Fig.~\ref{fig:method}) has a major axis in the NIRCam image that is well-aligned with the microshutter, and we observe a strong velocity gradient in the 2D spectrum with only a small offset between the major axis PA from the imaging and the median of the posterior distribution of the PA (shown in Fig.~\ref{fig:corner}). A prolate morphology is therefore highly unlikely for this object. The nature of the second object (JADES-NS-100016374) is more uncertain, as we only marginally detect rotation. If the kinematic major axis of the ionised gas differs strongly from the photometric major axis, the true rotational velocity and $v/\vdisp$ may be substantially higher than inferred with our modelling. }

\subsection{Comparing dynamical and stellar masses}\label{sec:mdyn}

We use the dynamical models to examine the mass budget of the galaxies. For a system in virial equilibrium, the dynamical mass enclosed within radius $r$ is computed as
\begin{equation}
    \Mdyn(<r) = k \frac{r \vc^2(r)}{G}\,,
\end{equation}
where $\vc$ is the circular velocity, $k$ the virial coefficient, and $G$ the gravitational constant. However, for comparison with the total stellar mass, we define a `total' dynamical mass as described in \citet{Price2022}:
\begin{equation}
    \Mdyn \equiv k_{\rm tot} \frac{\re \vc^2(\re) }{G}\,.
    \label{eq:mdyn}
\end{equation}
As we assume a thin disc model, we adopt $k_{\rm tot}=1.8$, which is the virial coefficient for an oblate potential with $q=0.2$ and $n\sim1-4$ \citep[][Fig. 4]{Price2022}. The true shape of the potential is not well-constrained however, and this choice for $k_{\rm tot}$ therefore introduces a systematic uncertainty in the dynamical mass estimates, as $k_{\rm tot}$ can vary by up to a factor two.

Following \citet{Burkert2010}, we compute the circular velocity as
\begin{equation}
    \vc(r) = \sqrt{ v^2(r) + 2 (r/r_{\rm d}) \vdisp^2  }\,,
\end{equation}
which accounts for the effects of pressure gradients on the rotational velocity, and depends on the disk scale length ($r_{\rm d}$). At the effective radius the pressure correction term reduces to $2(\re/r_{\rm d})=3.35$. {We note that for the one object with uncertain oblate/prolate morphology (Section~\ref{sec:results_vsigma}), this calculation of $\vc$ may be incorrect. However, as discussed in Section~\ref{sec:mass_budget}, the inferred dynamical mass is likely less affected.}

Next, we compare these total dynamical masses to stellar masses. To estimate stellar masses and SFRs, we perform SED modelling with the Bayesian fitting code BEAGLE \citep{Chevallard2016} to the low-resolution prism spectra. 
The fits were run adopting a two-component star formation history consisting of a delayed exponential with current burst, a \citet{Chabrier2003} initial mass function with an upper mass limit of $100\,\Msun$, and a \citet{Charlot2000} dust attenuation law assuming 40\% of the dust in the diffuse ISM. 
We note that the 1D prism spectra were flux-calibrated assuming a point-like morphology and without considering NIRCam photometry. Although this slit loss correction approximately corrects for the variation in the PSF FWHM with wavelength, there is a systematic offset between the total flux of the object and the flux captured by the slit. We estimate this aperture correction using our modelling software and the morphology in the long wavelength filter (F444W; measured with \texttt{lenstronomy} as described in Section~\ref{sec:fitting_description}), finding correction factors in the range $1.2-2.5$, and apply this to the stellar masses and SFRs. The inferred properties are presented in Appendix~\ref{sec:beagle}, together with an example prism spectrum and SED model.

\begin{figure}
    \centering
    \includegraphics[width=\linewidth]{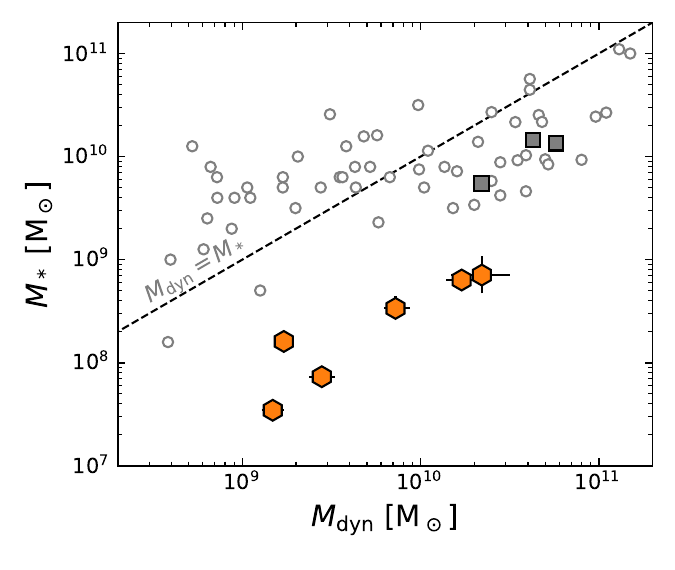}
    \caption{Stellar mass versus dynamical mass (Eq.~\ref{eq:mdyn}) as inferred from the prism and high-resolution spectroscopy, respectively. The dashed line shows the one-to-one relation between the two masses. Data points from the literature (circles, squares) are as described in Fig.~\ref{fig:vdisp}. As is to be expected, the dynamical masses are larger than the stellar masses for all objects in our sample. Surprisingly, however, the dynamical masses are substantially larger (up to a factor $\approx40$), most likely indicating large gas masses or large systematic uncertainties in the stellar mass estimates.}
    \label{fig:mdyn}
\end{figure}

We compare the estimated dynamical and stellar masses in Fig.~\ref{fig:mdyn}, and for reference plot the same ground-based near-infrared studies as in Figs.~\ref{fig:vdisp} and \ref{fig:vsigma}. As may be expected from the fact that $\Mdyn$ includes dark and baryonic mass, all objects in our sample have dynamical masses that are greater than the estimated stellar masses. However, the difference between the two masses is much larger than in previous studies, on average deviating by as much as a factor 30. {Only \citet{Topping2022} have reported similarly large stellar-to-dynamical mass discrepancies at $z\sim7$, albeit for more massive galaxies and based on spatially-integrated line width measurements instead of spatially-resolved dynamical modelling as in this paper. We discuss the possible origins of this discrepancy between the stellar and dynamical masses in detail in Section~\ref{sec:mass_budget}. }

\begin{table*}
    \caption{Dynamical modelling results: morphological properties and wavelength. Values are the median of the posterior probability distributions, and uncertainties reflect the $16^{\rm th}$ and $84^{\rm th}$ percentiles. }
    \centering
    \renewcommand{\arraystretch}{1.3}
   
    \begin{tabular}{l l l l l l l l l }
    \hline \hline
       ID  & $\lambda$  & flux $\times 10^{19}$ & dx & dy & $\re$  & $n$ & $q$ & PA  \\
        & ($\AA$) & ($\rm erg\,s^{-1}\,\AA^{-1}$) & (pitch)\tablefootmark{a} & (pitch)\tablefootmark{a} & (arcsec) &  & & (deg)  \\
       \hline
        JADES-NS-00016745 & $43104.8_{-0.7}^{+0.7}$ & $51_{-4}^{+4}$ & $-0.30_{-0.04}^{+0.04}$ & $0.014_{-0.018}^{+0.019}$ & $0.32_{-0.03}^{+0.03}$ & $1.41_{-0.13}^{+0.14}$ & $0.27_{-0.03}^{+0.03}$ & $67_{-3}^{+3}$   \\
        JADES-NS-00019606 & $34505.7_{-0.4}^{+0.3}$ & $30_{-2}^{+2}$ & $0.34_{-0.03}^{+0.03}$ & $0.025_{-0.013}^{+0.013}$ & $0.117_{-0.011}^{+0.013}$ & $0.25_{-0.04}^{+0.08}$ & $0.29_{-0.06}^{+0.06}$ & $130_{-4}^{+4}$ \\
        JADES-NS-00022251 & $34051.6_{-0.4}^{+0.4}$ & $77_{-4}^{+4}$ & $0.05_{-0.05}^{+0.05}$ & $-0.329_{-0.011}^{+0.010}$ & $0.122_{-0.006}^{+0.006}$ & $0.65_{-0.08}^{+0.09}$ & $0.46_{-0.08}^{+0.08}$ & $105_{-4}^{+4}$ \\
        JADES-NS-00047100\tablefootmark{b} & $42228.1_{-0.7}^{+0.8}$ & $44_{-3}^{+3}$ & $0.24_{-0.04}^{+0.04}$ & $0.203_{-0.014}^{+0.014}$ & $0.130_{-0.010}^{+0.011}$ & $2.0_{-0.7}^{+0.7}$ & $0.65_{-0.12}^{+0.11}$ & $65_{-8}^{+7}$  \\
        JADES-NS-10016374 & $42697.5_{-0.5}^{+0.5}$ & $16.9_{-1.0}^{+1.1}$ & $0.05_{-0.04}^{+0.04}$ & $0.004_{-0.008}^{+0.008}$ & $0.084_{-0.012}^{+0.013}$ & $1.4_{-0.5}^{+0.6}$ & $0.21_{-0.08}^{+0.08}$ & $130_{-6}^{+6}$\\
        JADES-NS-20086025 & $41381.8_{-1.1}^{+1.2}$ & $55_{-11}^{+14}$ & $0.03_{-0.08}^{+0.05}$ & $0.41_{-0.03}^{+0.03}$ & $0.36_{-0.10}^{+0.14}$ & $4.8_{-1.1}^{+0.8}$ & $0.27_{-0.10}^{+0.14}$ & $97_{-7}^{+7}$\\
    \hline  
    \end{tabular}
    \tablefoot{
    \tablefoottext{a}{The intrashutter offsets are measured in terms of shutter pitch rather than arcsec, as this unit is constant for all shutters and does not depend on the spatial distortion across the MSA.}
    \tablefoottext{b}{The morphology of this source is also discussed in \citet{Baker2023}.}
    }
    \label{tab:fit_results}
\end{table*}

\begin{table*}
    \caption{Dynamical modelling results: dynamical properties and model-derived quantities. Values are the median of the posterior probability distributions, and uncertainties reflect the $16^{\rm th}$ and $84^{\rm th}$ percentiles.  }
    \centering
    \renewcommand{\arraystretch}{1.3}
    \begin{tabular}{l l l l l l l l }
    \hline \hline
       ID & $z$ & $\rt/\re$ & $\va$ & $\sigma_0$ & $v(\re)$ & $v(\re)/\sigma_0$ & $\log(\Mdyn/\Msun)$ \\
         & & & ($\kms$) & ($\kms$)& ($\kms$) & &  \\ 
        \hline
        JADES-NS-00016745 & $5.56616_{-0.00011}^{+0.00010}$ & $0.7_{-0.3}^{+0.2}$ & $173_{-47}^{+43}$ & $55_{-2}^{+2}$ & $105_{-13}^{+13}$ & $1.9_{-0.2}^{+0.3}$ & $10.23_{-0.08}^{+0.08}$  \\
        JADES-NS-00019606 & $5.88979_{-0.00008}^{+0.00007}$ & $0.6_{-0.4}^{+0.3}$ & $0_{-11}^{+10}$ & $39.1_{-1.8}^{+1.8}$ & $5_{-3}^{+5}$ & $0.13_{-0.09}^{+0.14}$ & $9.17_{-0.06}^{+0.06}$ \\
        JADES-NS-00022251 & $5.79912_{-0.0007}^{+0.00007}$ & $0.75_{-0.26}^{+0.18}$ & $37_{-9}^{+9}$ & $39.0_{-1.0}^{+1.0}$ & $23_{-4}^{+4}$ & $0.58_{-0.11}^{+0.11}$ & $9.23_{-0.03}^{+0.03}$ \\
        JADES-NS-00047100\tablefootmark & $7.43173_{-0.00014}^{+0.00015}$ & $0.5_{-0.3}^{+0.3}$ & $132_{-36}^{+46}$ & $71_{-4}^{+4}$ & $91_{-16}^{+21}$ & $1.3_{-0.3}^{+0.3}$ & $9.85_{-0.07}^{+0.07}$  \\
        JADES-NS-10016374 & $5.50411_{-0.00007}^{+0.00007}$ & $0.6_{-0.4}^{+0.3}$ & $-23_{-15}^{+14}$\tablefootmark{a} & $62_{-2}^{+2}$ & $16_{-9}^{+10}$ & $0.26_{-0.14}^{+0.16}$ & $9.45_{-0.07}^{+0.07}$ \\
        JADES-NS-20086025 & $7.2627_{-0.0002}^{+0.0002}$ & $0.15_{-0.06}^{+0.10}$ & $-174_{-26}^{+22}$\tablefootmark{a} & $25_{-8}^{+6}$ & $155_{-17}^{+19}$ & $6.0_{-1.4}^{+2.9}$ & $10.31_{-0.19}^{+0.20}$ \\
    \hline
    \end{tabular}
    \tablefoot{
    \tablefoottext{a}{The sign of the model parameter $v_{\rm a}$ indicates the observed direction of the velocity gradient along the slit.}
    }
    \label{tab:fit_results2}
\end{table*}

\section{Discussion} \label{sec:discussion}

These data and modelling have taken us to a very new regime of galaxy kinematics: low-mass galaxies at $z>5$. Our modelling results are formally very well constrained, and the resulting parameter constraints (e.g., $M_*$ vs. $\Mdyn$) at face value imply spectacular results. Yet, a look at Fig.~\ref{fig:sample_overview} also makes it clear that our simple symmetric models may not capture the complex geometry of the systems. Therefore, our results warrant and require careful discussion.

\subsection{Clumpy cold discs or mergers?} \label{sec:mergers}

Using a forward modelling approach we have been able to separately constrain the morphology, velocity gradient and intrinsic velocity dispersion for each JADES object. To do so, we assumed an underlying model of a thin rotating disc  (Section~\ref{sec:model_description}). The velocity gradients measured in the context of our model suggest that the systems are dynamically relatively cold with higher than anticipated $v/\sigma$ ratios (Fig.~\ref{fig:vsigma}) based on an extrapolation of kinematic studies at $z<4$.

However, both observations and theoretical models have suggested that the rate of (major) mergers rises rapidly toward $z\sim6$ \citep[e.g.,][]{Rodriguez2015,Bowler2017,Duncan2019,OLeary2021}. It is therefore likely that some objects in our sample are merging systems, or have recently merged with another galaxy. We indeed find complex (emission line) morphologies for several objects, most notably apparent in objects JADES-NS-00016745 and JADES-NS-20086025 (see Fig.~\ref{fig:sample_overview}), and \citet{Baker2023} show that object JADES-NS-00047100 can be described by three separate morphological components. It is therefore possible that the rotational velocities inferred under the assumption of a virialised system may actually reflect the velocity offset between two (or more) objects, or velocity gradients induced by the gravitational interaction in a pre- or post-merger phase. 

Observations, on the other hand, also show that high-redshift galaxies often contain large star-forming clumps, and that the overall clumpiness of galaxies increases toward higher redshifts and lower masses \citep[e.g.,][]{Guo2015,Carniani2018,Zhang2019,Sattari2023}. Importantly, these clumps do not necessarily lead to a globally unstable system, and can be sustained within a rotationally-supported, albeit warm, disc \citep{FSchreiber2011,Mandelker2014}.

With the small angular scales ($\sim0.2\arcsec$) and velocity differences ($\Delta v\sim 200\,\kms$) involved for the systems studied in this paper, it is very difficult to distinguish between a merging system and star-forming clumps with ordered rotation. This degeneracy has been discussed extensively in the literature, although for lower redshifts and larger angular scales \citep[e.g.,][]{Krajnovic2006,Shapiro2008,Wisnioski2015,Rodrigues2017}. \citet{Simons2019} used simulations of merging galaxies to construct mock observations and hence quantify the frequency with which these systems are misclassified as rotating discs, showing that misclassifications are very common ($\approx50\%$), unless very stringent disc selection criteria are applied. Similarly, \citet{Hung2015,Hung2016} demonstrated that it becomes increasingly difficult to distinguish mergers from rotating systems toward later stages in the interaction between galaxies. On the other hand, \citet{Robertson2006} used hydrodynamical simulations to show that mergers between gas-rich systems can also lead to the formation of rotating discs with high angular momentum. Gravitational interaction between galaxies and the formation of rotating discs are thus not necessarily mutually exclusive, and the high gas fractions inferred in the next Section (\ref{sec:mass_budget}) are at face value consistent with the scenario proposed in these simulations.

Therefore, for any individual galaxy in our sample we cannot definitively conclude whether it is truly a rotating disc or an ongoing merging system. Although the NIRSpec MOS data are unprecedented in depth, resolution and sensitivity for galaxies at this mass and redshift, the objects are resolved by only a few resolution elements along a single spatial direction. Follow-up observations with the NIRSpec IFS mode can provide resolved 2D velocity field maps for these systems, which may then be compared against the disc selection criteria of \citet{Wisnioski2015} and \citet{Simons2019} to improve the constraints on their dynamical states.
However, the high-resolution NIRSpec IFS observations needed are not feasible for large samples of objects. It therefore seems inevitable at present to accept the fact that the nature of individual galaxies remains ambiguous. A statistical framework combining merger rates of galaxies and their observed emission line kinematics may provide a way forward to observationally constrain the settling of galaxies into cold discs at high redshifts, with number statistics that will be provided by upcoming surveys.

\subsection{Uncertainties in the mass budget}\label{sec:mass_budget}

We found a large discrepancy between the stellar and dynamical masses for the JADES objects. The dynamical mass presumably reflects the sum of the dark, stellar and gas mass within the effective radius. It is therefore not unexpected that the dynamical masses are larger than the stellar masses. However, the magnitude of the mass discrepancy -- more than an order of magnitude -- is surprising, as it is significantly larger than previous studies at lower redshifts at low mass \citep[e.g.,][]{Maseda2013}.

Considering the discussion of the previous section, we should first examine the robustness of our dynamical masses. To calculate the dynamical mass (Eq.~\ref{eq:mdyn}) we assumed that the galaxies are virialised; to compute the circular velocity, we assumed that the mass profile is approximately consistent with a rotating disc and exponential mass distribution. For the dispersion-dominated objects, the latter assumption may be problematic. If we instead assume a spherical mass distribution for these objects, we can instead follow the dynamical mass calculation for dispersion-supported systems by \citet{Cappellari2006}:
\begin{equation}
    \Mdyn = K(n)\frac{\re\sigma_0^2}{G}\,,
    \label{eq:mdyn_capp}
\end{equation}
where the virial coefficient depends on the S\'ersic index, $K(n) = 8.87 - 0.831n + 0.0241n^2$. However, for the low S\'ersic indices of our sample $K(n\approx1)\approx 8$ and is therefore comparable to the coefficients that enter Eq.~\ref{eq:mdyn}, as $k_{\rm tot}\vc^2 \approx 3.35 k_{\rm tot} \vdisp^2 \approx 7\vdisp^2$. {Similarly, in case of a prolate mass distribution we would expect $k_{\rm tot}\sim 4$ \citep{Price2022} and $\vc\approx\sqrt{3}\vdisp$}. In other words, the dynamical mass would not be overestimated by much if systems were actually dispersion dominated. {On the contrary, for the object in our sample with uncertain oblate/prolate morphology (Section~\ref{sec:results_vsigma}) the rotational velocity is possibly underestimated, which would lead to an underestimation of the dynamical mass and stellar-to-dynamical mass discrepancy. }

On the other hand, for the rotation-dominated objects $\Mdyn$ will be dominated by the value of $v(\re)$. If this velocity does not reflect a rotational velocity of a virialised system, but a velocity offset between two objects, then we cannot expect the dynamical mass estimate to be accurate. Both \citet{Simons2019} and \citet{Kohandel2019} explored the effects of incorrect physical and observational assumptions on the resulting $\vc$ and $\Mdyn$ estimates using mock observations of simulated galaxies. \citet{Simons2019} showed that for a merging system (noting that this is only a single simulation), the circular velocity is on average overestimated by a factor $\approx 1.5$ ($\approx 0.15\,$dex), which translates into an overestimation of $\Mdyn$ by a factor 2 ($0.3\,$dex). \citet{Kohandel2019} showed that, depending on the assumed inclination angle, $\Mdyn$ can be both under- and overestimated in the case of a merger, and find a mass discrepancy of $\approx \pm0.3\,$dex for velocity offsets of the same magnitude as found in our sample. Together with the uncertainty in the virial coefficient (see Section~\ref{sec:mdyn}), we therefore conclude that the dynamical masses may be overestimated by up to $\approx 0.3-0.6\,$dex, which cannot explain the large differences we find between the stellar and dynamical masses.

Yet, in the above we assume that the inferred gas kinematics are dominated by gravitational motions. If the velocity dispersions or velocity gradients are instead the result of non-gravitational motions, i.e. turbulence and outflows due to stellar feedback, then the dynamical masses may be severely overestimated. 
As is discussed in great detail in \citet{Uebler2019}, based on theoretical models the turbulence due to stellar feedback appears to be in the range of $\sim 10-20\,\kms$ \citep{Ostriker2011,Shetty2012,Krumholz2018}. This is significantly lower than the circular velocities measured for our sample and therefore cannot lead to a large bias in our dynamical masses. 

Outflows, however, may form a larger source of uncertainty. We have selected against objects in JADES with outflows as presented in \citet{Carniani2023}, who measured
outflow velocities $v_{\rm out}>200\,\kms$. Lower outflow velocities are difficult to detect, but may still be present in our data. We therefore turn to observations of starburst galaxies at low redshift for comparison. \citet{Heckman2015} and \citet{Xu2022} detected outflows using UV metal absorption lines, and demonstrated that the ratio $v_{\rm out}/\vc$ correlates with both the specific SFR (${\rm sSFR}\equiv{\rm SFR}/M_*$) and the SFR surface density ($\Sigma_{\rm SFR}$). Based on Fig.~10 of \citet{Xu2022} and the fact that for our sample $\rm sSFR\sim 10^{-8}\,yr^{-1}$ and $\Sigma_{\rm SFR}\sim 10\,\Msun\,yr^{-1}\,kpc^{-2}$, we estimate $v_{\rm out}/\vc \sim 3$. This would imply an overestimation of the circular velocity by a factor 3 or a factor 10 in the dynamical mass. However, it is unclear whether the outflowing gas traced by the rest-frame UV absorption lines is also traced by the rest-frame optical emission lines, and how this in turn would translate into the uncertainty on the dynamical mass. For example, \citet{Erb2006} found no correlation between the H$\alpha$ line width (and hence the dynamical mass) and galactic outflow velocities measured from rest-frame UV absorption lines for more massive star-forming galaxies at $z\approx2$.

\begin{figure}
    \centering
    \includegraphics[width=\linewidth]{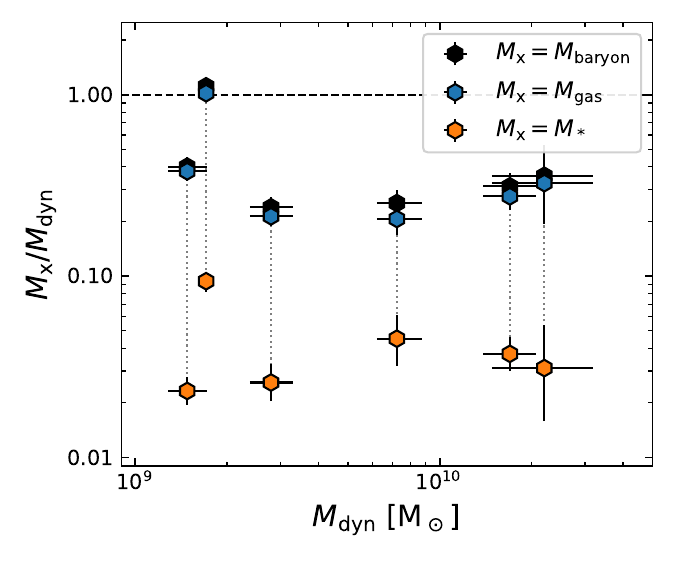}
    \caption{The baryonic, stellar and gas masses \citep[estimated using the relation between $\Sigma_{\rm SFR}$ and $\Sigma_{\rm gas}$ from][]{Kennicutt1998} as a fraction of the dynamical mass. We find that the gas mass, and hence the baryonic mass, is approximately one order of magnitude larger than the stellar mass. Although the inclusion of the gas component reduces the large discrepancy in mass found in Fig.~\ref{fig:mdyn}, a factor $3-4$ difference between the dynamical and baryonic mass still remains for all but one object. }
    \label{fig:mbar}
\end{figure}

If the dynamical mass is robust (at the factor 2 level), we should turn our attention to the other mass components that contribute to the total mass budget. An obvious component not discussed so far is the gas mass. Both observational and theoretical studies have shown that the gas content is important to take into account \citep[][]{Price2016,Wuyts2016}, as the typical gas fraction rises rapidly toward higher redshift and lower masses \citep[for a review, see][]{Tacconi2020}. We estimate the gas masses of our sample based on the SFRs obtained from the SED modelling to the prism spectroscopy. We use the inverse of the \citet{Kennicutt1998} relation between gas surface mass density ($\Sigma_{\rm gas}$) and SFR to infer the total gas mass. Although calibrated only at low redshifts, this likely provides a reasonable order-of-magnitude estimation for the high SFR surface densities of our sample \citep{Daddi2010,Kennicutt2012}. 

From this, we find gas masses of the order $\Mgas \sim 10^9\,\Msun$ with the average ratio $\langle\Mgas/M_*\rangle \approx 10$. We can now compare these gas masses and resulting baryonic masses ($\Mbar \equiv M_*+\Mgas$) to the dynamical masses in Fig.~\ref{fig:mbar}. Although the gas mass is large compared to the stellar mass ($f_{\rm gas} \equiv \Mgas/\Mbar \approx 0.90$; {consistent with measurements of more massive galaxies at $z\sim7$ from \citealt{Heintz2022}}), a discrepancy of approximately a factor $3-4$ between the baryonic and dynamical mass remains for all but one object. Our estimate of the gas masses carries a systematic uncertainty, as it for instance assumes a constant star formation efficiency. 
\citet{Price2020} show that a different estimator for the gas mass \citep[following][]{Tacconi2018} results in a mean gas mass difference of only $0.13\,$dex for a sample of more massive galaxies at cosmic noon. The same gas scaling relations are poorly, if at all, constrained at $z\sim6$ in the stellar mass regime considered in this work, and we therefore cannot make the same comparison for our sample. To increase the gas masses and make the baryonic masses consistent with the dynamical masses would, however, require a factor $\approx3$ decrease in the star formation efficiency for the majority of our sample. Yet, \citet{Pillepich2019} show that in the cosmological simulation TNG50 the gas fraction in low-mass galaxies ($M_*\sim10^9$) increases rapidly with redshift up to $z\sim3-4$, but then appears to flatten at higher redshifts with $\Mgas/\Mdyn \approx 0.2$. Further observations as well as simulated data at even lower stellar masses and higher redshifts will be necessary to constrain the gas masses for objects similar to the JADES sample.

Importantly, also the stellar masses may suffer from systematic uncertainties. The low-resolution spectroscopy spans a broad wavelength range ($1-5\,\micron$), probing the rest-frame UV to optical for all objects, and therefore provides good constraints on the recent star formation history (SFH). However, these measurements may suffer from an `outshining effect' in which a young star-forming population dominates the SED, making it near-impossible to detect an underlying population of older stars at rest-frame optical wavelengths \citep[e.g.,][]{Maraston2010,Pforr2012,Sorba2018,Gimenez2023,Tacchella2023,Whitler2023}. This is especially relevant for our sample, as we selected bright emission lines to perform our dynamical modelling, and these lines tend to have high equivalent widths. Using mock observations of cosmological simulations, \citet{Narayanan2023} show that the outshining effect may underestimate the stellar masses by $0.1-1.0$\,dex at $z\approx 7$ depending on the selected prior for the SFH. 
This makes the possible severe underestimate of the stellar mass the most problematic potential source of systematic errors in our mass budget. Imaging at longer wavelengths with JWST/MIRI and spatially-resolved SED fitting may offer improvements in the stellar mass estimates in the future \citep[e.g.,][]{Abdurrouf2023,Gimenez2023,Perez2023}. We note, however, that even this effect would play little role in the regime where $\Mdyn$ is $30\times M_*$.

In summary, a number of systematic effects may be contribute to the large discrepancy of a factor 30 between the stellar and dynamical masses for this low-mass, high-redshift sample. We argue that the dynamical masses are relatively well-constrained, with at most a $0.3-0.6\,$dex uncertainty even in the case of an ongoing merger, although we cannot rule out a bias in some of the dynamical mass measurements due to galactic-scale outflows. Clearly, a substantial amount of gas must be present in these highly star-forming galaxies, and we estimate this can account for $\approx1\,$dex in the stellar-to-dynamical mass discrepancy. Although the gas masses are uncertain, and scaling relations between SFR and gas surface densities have significant scatter, we believe it is unlikely that the gas masses are underestimated by a large factor for all objects in our sample. This leaves the possibility that instead the stellar masses may be significantly underestimated, as we lack constraints at longer rest-frame wavelengths where an older stellar population may be measurable.

In this discussion we have neglected one mass component thus far: the dark matter. On the small spatial scales probed ($\approx 1\,$kpc) this may not appear to be a dominant factor in the mass budget, particularly as multiple studies have shown a rapid increase in the central baryon fraction toward higher redshifts \citep[e.g.,][]{vDokkum2015,Price2016,Wuyts2016,Genzel2017,Genzel2020}. Yet, it is interesting to consider a situation where there is significant dark matter within the effective radii at these redshifts. Within the $\Lambda$CDM cosmological model, dark matter dominates the mass content of the Universe. Under hierarchical structure formation, small dark matter haloes form first, whereas stars only form after the gas within those haloes has cooled sufficiently, subsequently growing into galaxies through accretion \citep[cold gas streams, mergers; e.g.,][]{White1978,Dekel2009a,Oser2010}. In this scenario, it may be possible that very young galaxies are dark matter dominated even in the central regions, as the baryonic mass is still under assembly. {This is of particular interest, as galaxies in the stellar mass regime probed in this paper ($\sim10^8\,\Msun$) may be the progenitors of galaxies of $M_*\sim10^{11}\,\Msun$ at $z=0$ \citep{Moster2018,Behroozi2019}, which have baryon-dominated centres. As the aforementioned observations at cosmic noon are typically also of more massive galaxies ($M_*\sim10^{10-11}\,\Msun$), the difference in the central baryon fractions between our work and measurements at cosmic noon may therefore not be in tension, but reflect the time evolution in the distribution of dark and baryonic mass as galaxies grow and assemble their stellar mass. We explore this idea further using cosmological simulations in de Graaff et al. (in prep.). However, to definitively conclude whether this scenario may apply to the JADES galaxies will require a thorough understanding of the systematic uncertainties on the different mass components. }

\section{Conclusions}\label{sec:conclusion}

We use the JADES spectroscopic sample in GOODS-S to select six targets at $z=5.5-7.4$ that are spatially extended in NIRCam imaging, and for which high-resolution ($R\sim2700$) spectroscopy was obtained with the NIRSpec MSA. We show that these galaxies lie in a previously unprobed part of parameter space: not only because of their high redshifts, but also their small sizes ($\sim 1\,$kpc) and low stellar masses ($M_*\sim10^8\,\Msun$). 
The high-resolution spectra reveal rest-frame optical emission lines ([O{\sc iii}] and H$\alpha$) that are broadened and have spatial velocity gradients. 

To extract the dynamical properties we model the objects as thin, but warm, rotating discs. We describe a novel forward modelling software to account for several complexities of data taken with the NIRSpec instrument: the PSF, shutter geometry and bar shadows, and pixellation. Using NIRCam imaging as a prior on the emission line morphology, we are able to constrain the rotational velocities and velocity dispersions of the objects in our sample, and hence also estimate dynamical masses. Our findings can be summarised as follows. 

\begin{itemize}
    \item The objects in our sample are small ($\re\sim 0.5 - 2\,$kpc), of low stellar mass ($M_*\sim 10^{7.5-8.9}$) and modest star formation rates ($\rm SFR\sim 2-20\,\Msun\,yr^{-1}$), which we infer from SED modelling to low-resolution NIRSpec spectroscopy. The gas masses implied by the SFRs are on average $10\times$ larger than the stellar masses.
    \item We find intrinsic velocity dispersions in the range $\vdisp\approx 30-70\,\kms$, which is consistent with studies reporting the velocity dispersions of more massive galaxies at cosmic noon. 
    \item Three out of six objects show significant spatial velocity gradients, resulting in $v/\sigma\approx 1-6$. Under the assumption of our thin disc model, this implies that the high-redshift objects are rotation-dominated discs. However, we cannot rule out the possibility that the detected velocity gradients reflect velocity offsets between interacting galaxies.
    \item Comparison between the dynamical and stellar masses reveals a surprising discrepancy of a factor $10-40$. 
    After accounting for the large gas masses, the dynamical masses still remain larger than the baryonic masses by a factor $\sim3$. 
    \item We argue that the dynamical masses are robust within a factor $2-4$ even in the case of an ongoing merger. Only the presence of outflows, if these were to dominate the observed emission line kinematics, can substantially lower the inferred dynamical masses. However, the baryonic-to-dynamical mass discrepancy might also imply that the centres of these objects are dark-matter dominated. Moreover, there are large systematic uncertainties on the stellar and gas masses. The baryonic masses can be reconciled with the dynamical masses if the star formation efficiency in these objects is a factor 3 lower than initially assumed.

\end{itemize}

Our work provides a first demonstration of the powerful capabilities of the NIRSpec MOS mode to perform spatially- and spectrally-resolved analyses. Crucially, this enables the study of galaxy kinematics in a highly efficient manner, as a single observation can probe a wide range in redshift, mass and SFR. With larger spectroscopic samples using the high-resolution MOS mode currently being acquired, JWST NIRSpec will in the near-future allow for statistical analyses of the origins and settling of disc galaxies in the early Universe.

\begin{acknowledgements}
    AdG thanks M. Fouesneau and I. Momcheva for their valuable feedback during the development of \texttt{msafit}, and J. Davies for helpful discussions on JWST and NIRSpec data. This research makes use of ESA Datalabs (datalabs.esa.int), an initiative by ESA’s Data Science and Archives Division in the Science and Operations Department, Directorate of Science. This research made use of POPPY, an open-source optical propagation python package originally developed for the James Webb Space Telescope project \citep{Perrin2012}.

SC acknowledges support by European Union’s HE ERC Starting Grant No. 101040227 - WINGS. ECL acknowledges support of an STFC Webb Fellowship (ST/W001438/1)
RM and WB acknowledge support by the Science and Technology Facilities Council (STFC), by the ERC through Advanced Grant 695671 “QUENCH”. RM also acknowledges support by the UKRI Frontier Research grant RISEandFALL and funding from a research professorship from the Royal Society.
 AJB, AJC, JC \& GCJ acknowledge funding from the "FirstGalaxies" Advanced Grant from the European Research Council (ERC) under the European Union’s Horizon 2020 research and innovation programme (Grant agreement No. 789056)
 SA acknowledges support from Grant PID2021-127718NB-I00 funded by the Spanish Ministry of Science and Innovation/State Agency of Research (MICIN/AEI/ 10.13039/501100011033). 
H{\"U} gratefully acknowledges support by the Isaac Newton Trust and by the Kavli Foundation through a Newton-Kavli Junior Fellowship.
 DJE is supported as a Simons Investigator. DJE, BJJ and CNAW are supported by a JWST/NIRCam contract to the University of Arizona, NAS5-02015.
BER acknowledges support from the NIRCam Science Team contract to the University of Arizona, NAS5-02015. 
The research of CCW is supported by NOIRLab, which is managed by the Association of Universities for Research in Astronomy (AURA) under a cooperative agreement with the National Science Foundation.
KB acknowledges support from the Australian Research Council Centre of Excellence for All Sky Astrophysics in 3 Dimensions (ASTRO 3D), through project number CE170100013. RH acknowledges funding provided by the Johns Hopkins University, Institute for Data Intensive Engineering and Science (IDIES).
 
\end{acknowledgements}

\bibliographystyle{aa}
\bibliography{nirspec}

\begin{thebibliography}{118}
\expandafter\ifx\csname natexlab\endcsname\relax\def\natexlab#1{#1}\fi

\bibitem[{{Abdurro'uf} {et~al.}(2023){Abdurro'uf}, {Coe}, {Jung}, {Ferguson},
  {Brammer}, {Iyer}, {Bradley}, {Dayal}, {Windhorst}, {Zitrin}, {Meena},
  {Oguri}, {Diego}, {Kokorev}, {Dimauro}, {Adamo}, {Conselice}, {Welch},
  {Vanzella}, {Hsiao}, {Xu}, {Roy}, \& {Mulcahey}}]{Abdurrouf2023}
{Abdurro'uf}, {Coe}, D., {Jung}, I., {et~al.} 2023, \apj, 945, 117

\bibitem[{{Alves de Oliveira} {et~al.}(2022){Alves de Oliveira},
  {L{\"u}tzgendorf}, {Zeidler}, {Giardino}, {Ferruit}, {Kumari}, {Rawle},
  {Birkmann}, {B{\"o}ker}, {Proffitt}, {Sirianni}, \& {Te Plate}}]{Alves2022}
{Alves de Oliveira}, C., {L{\"u}tzgendorf}, N., {Zeidler}, P., {et~al.} 2022,
  in Society of Photo-Optical Instrumentation Engineers (SPIE) Conference
  Series, Vol. 12180, Space Telescopes and Instrumentation 2022: Optical,
  Infrared, and Millimeter Wave, ed. L.~E. {Coyle}, S.~{Matsuura}, \& M.~D.
  {Perrin}, 121803S

\bibitem[{{Baker} {et~al.}(2023){Baker}, {Tacchella}, {Johnson}, {Nelson},
  {Suess}, {D'Eugenio}, {Curti}, {de Graaff}, {Ji}, {Maiolino}, {Robertson},
  {Scholtz}, {Alberts}, {Arribas}, {Boyett}, {Bunker}, {Carniani}, {Charlot},
  {Chen}, {Chevallard}, {Curtis-Lake}, {Danhaive}, {DeCoursey}, {Egami},
  {Eisenstein}, {Endsley}, {Hausen}, {Helton}, {Kumari}, {Looser}, {Maseda},
  {Pusk{\'a}s}, {Rieke}, {Sandles}, {Sun}, {{\"U}bler}, {Williams}, {Willmer},
  \& {Witstok}}]{Baker2023}
{Baker}, W.~M., {Tacchella}, S., {Johnson}, B.~D., {et~al.} 2023, arXiv
  e-prints, arXiv:2306.02472

\bibitem[{{Behroozi} {et~al.}(2019){Behroozi}, {Wechsler}, {Hearin}, \&
  {Conroy}}]{Behroozi2019}
{Behroozi}, P., {Wechsler}, R.~H., {Hearin}, A.~P., \& {Conroy}, C. 2019,
  \mnras, 488, 3143

\bibitem[{{Birrer} \& {Amara}(2018)}]{Birrer2018}
{Birrer}, S. \& {Amara}, A. 2018, Physics of the Dark Universe, 22, 189

\bibitem[{{Birrer} {et~al.}(2021){Birrer}, {Shajib}, {Gilman}, {Galan},
  {Aalbers}, {Millon}, {Morgan}, {Pagano}, {Park}, {Teodori}, {Tessore},
  {Ueland}, {Van de Vyvere}, {Wagner-Carena}, {Wempe}, {Yang}, {Ding},
  {Schmidt}, {Sluse}, {Zhang}, \& {Amara}}]{Birrer2021}
{Birrer}, S., {Shajib}, A., {Gilman}, D., {et~al.} 2021, The Journal of Open
  Source Software, 6, 3283

\bibitem[{{B{\"o}ker} {et~al.}(2022){B{\"o}ker}, {Arribas}, {L{\"u}tzgendorf},
  {Alves de Oliveira}, {Beck}, {Birkmann}, {Bunker}, {Charlot}, {de Marchi},
  {Ferruit}, {Giardino}, {Jakobsen}, {Kumari}, {L{\'o}pez-Caniego}, {Maiolino},
  {Manjavacas}, {Marston}, {Moseley}, {Muzerolle}, {Ogle}, {Pirzkal},
  {Rauscher}, {Rawle}, {Rix}, {Sabbi}, {Sargent}, {Sirianni}, {te Plate},
  {Valenti}, {Willott}, \& {Zeidler}}]{Boeker2022}
{B{\"o}ker}, T., {Arribas}, S., {L{\"u}tzgendorf}, N., {et~al.} 2022, \aap,
  661, A82

\bibitem[{{B{\"o}ker} {et~al.}(2023){B{\"o}ker}, {Beck}, {Birkmann},
  {Giardino}, {Keyes}, {Kumari}, {Muzerolle}, {Rawle}, {Zeidler}, {Abul-Huda},
  {Alves de Oliveira}, {Arribas}, {Bechtold}, {Bhatawdekar}, {Bonaventura},
  {Bunker}, {Cameron}, {Carniani}, {Charlot}, {Curti}, {Espinoza}, {Ferruit},
  {Franx}, {Jakobsen}, {Karakla}, {L{\'o}pez-Caniego}, {L{\"u}tzgendorf},
  {Maiolino}, {Manjavacas}, {Marston}, {Moseley}, {Ogle}, {Perna},
  {Pe{\~n}a-Guerrero}, {Pirzkal}, {Plesha}, {Proffitt}, {Rauscher}, {Rix},
  {Rodr{\'\i}guez del Pino}, {Rustamkulov}, {Sabbi}, {Sing}, {Sirianni}, {te
  Plate}, {{\'U}beda}, {Wahlgren}, {Wislowski}, {Wu}, \&
  {Willott}}]{Boeker2023}
{B{\"o}ker}, T., {Beck}, T.~L., {Birkmann}, S.~M., {et~al.} 2023, \pasp, 135,
  038001

\bibitem[{{Bowler} {et~al.}(2017){Bowler}, {Dunlop}, {McLure}, \&
  {McLeod}}]{Bowler2017}
{Bowler}, R.~A.~A., {Dunlop}, J.~S., {McLure}, R.~J., \& {McLeod}, D.~J. 2017,
  \mnras, 466, 3612

\bibitem[{{Bunker} {et~al.}(2023){Bunker}, {Cameron}, {Curtis-Lake},
  {Jakobsen}, {Carniani}, {Curti}, {Witstok}, {Maiolino}, {D'Eugenio},
  {Looser}, {Willott}, {Bonaventura}, {Hainline}, {Uebler}, {Willmer},
  {Saxena}, {Smit}, {Alberts}, {Arribas}, {Baker}, {Baum}, {Bhatawdekar},
  {Bowler}, {Boyett}, {Charlot}, {Chen}, {Chevallard}, {Circosta}, {DeCoursey},
  {de Graaff}, {Egami}, {Eisenstein}, {Endsley}, {Ferruit}, {Giardino},
  {Hausen}, {Helton}, {Hviding}, {Ji}, {Johnson}, {Jones}, {Kumari}, {Laseter},
  {Luetzgendorf}, {Maseda}, {Nelson}, {Parlanti}, {Perna}, {Rawle}, {Rix},
  {Rieke}, {Robertson}, {Rodriguez Del Pino}, {Sandles}, {Scholtz}, {Sharpe},
  {Skarbinski}, {Stark}, {Sun}, {Tacchella}, {Topping}, {Villanueva},
  {Wallace}, {Williams}, \& {Woodrum}}]{Bunker2023}
{Bunker}, A.~J., {Cameron}, A.~J., {Curtis-Lake}, E., {et~al.} 2023, arXiv
  e-prints, arXiv:2306.02467

\bibitem[{{Burkert} {et~al.}(2010){Burkert}, {Genzel}, {Bouch{\'e}}, {Cresci},
  {Khochfar}, {Sommer-Larsen}, {Sternberg}, {Naab}, {F{\"o}rster Schreiber},
  {Tacconi}, {Shapiro}, {Hicks}, {Lutz}, {Davies}, {Buschkamp}, \&
  {Genel}}]{Burkert2010}
{Burkert}, A., {Genzel}, R., {Bouch{\'e}}, N., {et~al.} 2010, \apj, 725, 2324

\bibitem[{{Cappellari}(2016)}]{Cappellari2016}
{Cappellari}, M. 2016, \araa, 54, 597

\bibitem[{{Cappellari} {et~al.}(2006){Cappellari}, {Bacon}, {Bureau}, {Damen},
  {Davies}, {de Zeeuw}, {Emsellem}, {Falc{\'o}n-Barroso}, {Krajnovi{\'c}},
  {Kuntschner}, {McDermid}, {Peletier}, {Sarzi}, {van den Bosch}, \& {van de
  Ven}}]{Cappellari2006}
{Cappellari}, M., {Bacon}, R., {Bureau}, M., {et~al.} 2006, \mnras, 366, 1126

\bibitem[{{Carniani} {et~al.}(2018){Carniani}, {Maiolino}, {Amorin},
  {Pentericci}, {Pallottini}, {Ferrara}, {Willott}, {Smit}, {Matthee},
  {Sobral}, {Santini}, {Castellano}, {De Barros}, {Fontana}, {Grazian}, \&
  {Guaita}}]{Carniani2018}
{Carniani}, S., {Maiolino}, R., {Amorin}, R., {et~al.} 2018, \mnras, 478, 1170

\bibitem[{{Carniani} {et~al.}(2023){Carniani}, {Venturi}, {Parlanti}, {de
  Graaff}, {Maiolino}, {Arribas}, {Bonaventura}, {Boyett}, {Bunker}, {Cameron},
  {Charlot}, {Chevallard}, {Curti}, {Curtis-Lake}, {Eisenstein}, {Giardino},
  {Hausen}, {Kumari}, {Maseda}, {Nelson}, {Perna}, {Rix}, {Robertson},
  {Rodr{\'\i}guez Del Pino}, {Sandles}, {Scholtz}, {Simmonds}, {Smit},
  {Tacchella}, {{\"U}bler}, {Williams}, {Willott}, \& {Witstok}}]{Carniani2023}
{Carniani}, S., {Venturi}, G., {Parlanti}, E., {et~al.} 2023, arXiv e-prints,
  arXiv:2306.11801

\bibitem[{{Ceverino} {et~al.}(2012){Ceverino}, {Dekel}, {Mandelker},
  {Bournaud}, {Burkert}, {Genzel}, \& {Primack}}]{Ceverino2012}
{Ceverino}, D., {Dekel}, A., {Mandelker}, N., {et~al.} 2012, \mnras, 420, 3490

\bibitem[{{Chabrier}(2003)}]{Chabrier2003}
{Chabrier}, G. 2003, \apjl, 586, L133

\bibitem[{{Charlot} \& {Fall}(2000)}]{Charlot2000}
{Charlot}, S. \& {Fall}, S.~M. 2000, \apj, 539, 718

\bibitem[{{Chevallard} \& {Charlot}(2016)}]{Chevallard2016}
{Chevallard}, J. \& {Charlot}, S. 2016, \mnras, 462, 1415

\bibitem[{{Courteau}(1997)}]{Courteau1997}
{Courteau}, S. 1997, \aj, 114, 2402

\bibitem[{{Curtis-Lake} {et~al.}(2023){Curtis-Lake}, {Carniani}, {Cameron},
  {Charlot}, {Jakobsen}, {Maiolino}, {Bunker}, {Witstok}, {Smit}, {Chevallard},
  {Willott}, {Ferruit}, {Arribas}, {Bonaventura}, {Curti}, {D'Eugenio},
  {Franx}, {Giardino}, {Looser}, {L{\"u}tzgendorf}, {Maseda}, {Rawle}, {Rix},
  {Rodr{\'\i}guez del Pino}, {{\"U}bler}, {Sirianni}, {Dressler}, {Egami},
  {Eisenstein}, {Endsley}, {Hainline}, {Hausen}, {Johnson}, {Rieke},
  {Robertson}, {Shivaei}, {Stark}, {Tacchella}, {Williams}, {Willmer},
  {Bhatawdekar}, {Bowler}, {Boyett}, {Chen}, {de Graaff}, {Helton}, {Hviding},
  {Jones}, {Kumari}, {Lyu}, {Nelson}, {Perna}, {Sandles}, {Saxena}, {Suess},
  {Sun}, {Topping}, {Wallace}, \& {Whitler}}]{CurtisLake2023}
{Curtis-Lake}, E., {Carniani}, S., {Cameron}, A., {et~al.} 2023, Nature
  Astronomy, 7, 622

\bibitem[{{Daddi} {et~al.}(2010){Daddi}, {Elbaz}, {Walter}, {Bournaud},
  {Salmi}, {Carilli}, {Dannerbauer}, {Dickinson}, {Monaco}, \&
  {Riechers}}]{Daddi2010}
{Daddi}, E., {Elbaz}, D., {Walter}, F., {et~al.} 2010, \apjl, 714, L118

\bibitem[{{Dekel} {et~al.}(2009{\natexlab{a}}){Dekel}, {Birnboim}, {Engel},
  {Freundlich}, {Goerdt}, {Mumcuoglu}, {Neistein}, {Pichon}, {Teyssier}, \&
  {Zinger}}]{Dekel2009a}
{Dekel}, A., {Birnboim}, Y., {Engel}, G., {et~al.} 2009{\natexlab{a}}, \nat,
  457, 451

\bibitem[{{Dekel} {et~al.}(2009{\natexlab{b}}){Dekel}, {Sari}, \&
  {Ceverino}}]{Dekel2009b}
{Dekel}, A., {Sari}, R., \& {Ceverino}, D. 2009{\natexlab{b}}, \apj, 703, 785

\bibitem[{{Dorner} {et~al.}(2016){Dorner}, {Giardino}, {Ferruit}, {Alves de
  Oliveira}, {Birkmann}, {B{\"o}ker}, {De Marchi}, {Gnata}, {K{\"o}hler},
  {Sirianni}, \& {Jakobsen}}]{Dorner2016}
{Dorner}, B., {Giardino}, G., {Ferruit}, P., {et~al.} 2016, \aap, 592, A113

\bibitem[{{Duncan} {et~al.}(2019){Duncan}, {Conselice}, {Mundy}, {Bell},
  {Donley}, {Galametz}, {Guo}, {Grogin}, {Hathi}, {Kartaltepe}, {Kocevski},
  {Koekemoer}, {P{\'e}rez-Gonz{\'a}lez}, {Mantha}, {Snyder}, \&
  {Stefanon}}]{Duncan2019}
{Duncan}, K., {Conselice}, C.~J., {Mundy}, C., {et~al.} 2019, \apj, 876, 110

\bibitem[{{Eisenstein} {et~al.}(2023){Eisenstein}, {Willott}, {Alberts},
  {Arribas}, {Bonaventura}, {Bunker}, {Cameron}, {Carniani}, {Charlot},
  {Curtis-Lake}, {D'Eugenio}, {Endsley}, {Ferruit}, {Giardino}, {Hainline},
  {Hausen}, {Jakobsen}, {Johnson}, {Maiolino}, {Rieke}, {Rieke}, {Rix},
  {Robertson}, {Stark}, {Tacchella}, {Williams}, {Willmer}, {Baker}, {Baum},
  {Bhatawdekar}, {Boyett}, {Chen}, {Chevallard}, {Circosta}, {Curti},
  {Danhaive}, {DeCoursey}, {de Graaff}, {Dressler}, {Egami}, {Helton},
  {Hviding}, {Ji}, {Jones}, {Kumari}, {L{\"u}tzgendorf}, {Laseter}, {Looser},
  {Lyu}, {Maseda}, {Nelson}, {Parlanti}, {Perna}, {Pusk{\'a}s}, {Rawle},
  {Rodr{\'\i}guez Del Pino}, {Sandles}, {Saxena}, {Scholtz}, {Sharpe},
  {Shivaei}, {Silcock}, {Simmonds}, {Skarbinski}, {Smit}, {Stone}, {Suess},
  {Sun}, {Tang}, {Topping}, {{\"U}bler}, {Villanueva}, {Wallace}, {Whitler},
  {Witstok}, \& {Woodrum}}]{Eisenstein2023}
{Eisenstein}, D.~J., {Willott}, C., {Alberts}, S., {et~al.} 2023, arXiv
  e-prints, arXiv:2306.02465

\bibitem[{{Erb} {et~al.}(2006){Erb}, {Steidel}, {Shapley}, {Pettini}, {Reddy},
  \& {Adelberger}}]{Erb2006}
{Erb}, D.~K., {Steidel}, C.~C., {Shapley}, A.~E., {et~al.} 2006, \apj, 646, 107

\bibitem[{{Falc{\'o}n-Barroso} {et~al.}(2019){Falc{\'o}n-Barroso}, {van de
  Ven}, {Lyubenova}, {Mendez-Abreu}, {Aguerri}, {Garc{\'\i}a-Lorenzo},
  {Bekerait{\'e}}, {S{\'a}nchez}, {Husemann}, {Garc{\'\i}a-Benito},
  {Gonz{\'a}lez Delgado}, {Mast}, {Walcher}, {Zibetti}, {Zhu},
  {Barrera-Ballesteros}, {Galbany}, {S{\'a}nchez-Bl{\'a}zquez}, {Singh}, {van
  den Bosch}, {Wild}, {Bland-Hawthorn}, {Cid Fernandes}, {de
  Lorenzo-C{\'a}ceres}, {Gallazzi}, {Marino}, {M{\'a}rquez}, {Peletier},
  {P{\'e}rez}, {P{\'e}rez}, {Roth}, {Rosales-Ortega}, {Ruiz-Lara}, {Wisotzki},
  \& {Ziegler}}]{Falcon2019}
{Falc{\'o}n-Barroso}, J., {van de Ven}, G., {Lyubenova}, M., {et~al.} 2019,
  \aap, 632, A59

\bibitem[{{Ferruit} {et~al.}(2022){Ferruit}, {Jakobsen}, {Giardino}, {Rawle},
  {Alves de Oliveira}, {Arribas}, {Beck}, {Birkmann}, {B{\"o}ker}, {Bunker},
  {Charlot}, {de Marchi}, {Franx}, {Henry}, {Karakla}, {Kassin}, {Kumari},
  {L{\'o}pez-Caniego}, {L{\"u}tzgendorf}, {Maiolino}, {Manjavacas}, {Marston},
  {Moseley}, {Muzerolle}, {Pirzkal}, {Rauscher}, {Rix}, {Sabbi}, {Sirianni},
  {te Plate}, {Valenti}, {Willott}, \& {Zeidler}}]{Ferruit2022}
{Ferruit}, P., {Jakobsen}, P., {Giardino}, G., {et~al.} 2022, \aap, 661, A81

\bibitem[{{Foreman-Mackey} {et~al.}(2013){Foreman-Mackey}, {Hogg}, {Lang}, \&
  {Goodman}}]{emcee}
{Foreman-Mackey}, D., {Hogg}, D.~W., {Lang}, D., \& {Goodman}, J. 2013, \pasp,
  125, 306

\bibitem[{{F{\"o}rster Schreiber} {et~al.}(2018){F{\"o}rster Schreiber},
  {Renzini}, {Mancini}, {Genzel}, {Bouch{\'e}}, {Cresci}, {Hicks}, {Lilly},
  {Peng}, {Burkert}, {Carollo}, {Cimatti}, {Daddi}, {Davies}, {Genel}, {Kurk},
  {Lang}, {Lutz}, {Mainieri}, {McCracken}, {Mignoli}, {Naab}, {Oesch},
  {Pozzetti}, {Scodeggio}, {Shapiro Griffin}, {Shapley}, {Sternberg},
  {Tacchella}, {Tacconi}, {Wuyts}, \& {Zamorani}}]{SINS2018}
{F{\"o}rster Schreiber}, N.~M., {Renzini}, A., {Mancini}, C., {et~al.} 2018,
  \apjs, 238, 21

\bibitem[{{F{\"o}rster Schreiber} {et~al.}(2011){F{\"o}rster Schreiber},
  {Shapley}, {Erb}, {Genzel}, {Steidel}, {Bouch{\'e}}, {Cresci}, \&
  {Davies}}]{FSchreiber2011}
{F{\"o}rster Schreiber}, N.~M., {Shapley}, A.~E., {Erb}, D.~K., {et~al.} 2011,
  \apj, 731, 65

\bibitem[{{F{\"o}rster Schreiber} \& {Wuyts}(2020)}]{FSchreiber2020}
{F{\"o}rster Schreiber}, N.~M. \& {Wuyts}, S. 2020, \araa, 58, 661

\bibitem[{{Fraternali} {et~al.}(2021){Fraternali}, {Karim}, {Magnelli},
  {G{\'o}mez-Guijarro}, {Jim{\'e}nez-Andrade}, \& {Posses}}]{Fraternali2021}
{Fraternali}, F., {Karim}, A., {Magnelli}, B., {et~al.} 2021, \aap, 647, A194

\bibitem[{{Gardner} {et~al.}(2023){Gardner}, {Mather}, {Abbott}, {Abell},
  {Abernathy}, {Abney}, {Abraham}, {Abraham}, {Abul-Huda}, {Acton}, {Adams},
  {Adams}, {Adler}, {Adriaensen}, {Aguilar}, {Ahmed}, {Ahmed}, {Ahmed},
  {Albat}, {Albert}, {Alberts}, {Aldridge}, {Allen}, {Allen}, {Altenburg},
  {Altunc}, {Alvarez}, \& {{\'A}lvarez-M{\'a}rquez}}]{Gardner2023}
{Gardner}, J.~P., {Mather}, J.~C., {Abbott}, R., {et~al.} 2023, \pasp, 135,
  068001

\bibitem[{{Genel} {et~al.}(2012){Genel}, {Dekel}, \& {Cacciato}}]{Genel2012}
{Genel}, S., {Dekel}, A., \& {Cacciato}, M. 2012, \mnras, 425, 788

\bibitem[{{Genzel} {et~al.}(2017){Genzel}, {F{\"o}rster Schreiber},
  {{\"U}bler}, {Lang}, {Naab}, {Bender}, {Tacconi}, {Wisnioski}, {Wuyts},
  {Alexander}, {Beifiori}, {Belli}, {Brammer}, {Burkert}, {Carollo}, {Chan},
  {Davies}, {Fossati}, {Galametz}, {Genel}, {Gerhard}, {Lutz}, {Mendel},
  {Momcheva}, {Nelson}, {Renzini}, {Saglia}, {Sternberg}, {Tacchella},
  {Tadaki}, \& {Wilman}}]{Genzel2017}
{Genzel}, R., {F{\"o}rster Schreiber}, N.~M., {{\"U}bler}, H., {et~al.} 2017,
  \nat, 543, 397

\bibitem[{{Genzel} {et~al.}(2020){Genzel}, {Price}, {{\"U}bler}, {F{\"o}rster
  Schreiber}, {Shimizu}, {Tacconi}, {Bender}, {Burkert}, {Contursi}, {Coogan},
  {Davies}, {Davies}, {Dekel}, {Herrera-Camus}, {Lee}, {Lutz}, {Naab}, {Neri},
  {Nestor}, {Renzini}, {Saglia}, {Schuster}, {Sternberg}, {Wisnioski}, \&
  {Wuyts}}]{Genzel2020}
{Genzel}, R., {Price}, S.~H., {{\"U}bler}, H., {et~al.} 2020, \apj, 902, 98

\bibitem[{{Giardino} {et~al.}(2019){Giardino}, {Ferruit}, {Chevallard},
  {Curtis-Lake}, {Bonaventura}, {Jakobsen}, {Jarno}, {Pecontal}, \&
  {Piqueras}}]{Giardino2019}
{Giardino}, G., {Ferruit}, P., {Chevallard}, J., {et~al.} 2019, in Astronomical
  Society of the Pacific Conference Series, Vol. 523, Astronomical Data
  Analysis Software and Systems XXVII, ed. P.~J. {Teuben}, M.~W. {Pound}, B.~A.
  {Thomas}, \& E.~M. {Warner}, 645

\bibitem[{{Giardino} {et~al.}(2016){Giardino}, {Luetzgendorf}, {Ferruit},
  {Dorner}, {Alves de Oliveira}, {Birkmann}, {Boeker}, {Rawle}, \&
  {Sirianni}}]{Giardino2016}
{Giardino}, G., {Luetzgendorf}, N., {Ferruit}, P., {et~al.} 2016, in Society of
  Photo-Optical Instrumentation Engineers (SPIE) Conference Series, Vol. 9904,
  Space Telescopes and Instrumentation 2016: Optical, Infrared, and Millimeter
  Wave, ed. H.~A. {MacEwen}, G.~G. {Fazio}, M.~{Lystrup}, N.~{Batalha},
  N.~{Siegler}, \& E.~C. {Tong}, 990445

\bibitem[{{Gim{\'e}nez-Arteaga} {et~al.}(2023){Gim{\'e}nez-Arteaga}, {Oesch},
  {Brammer}, {Valentino}, {Mason}, {Weibel}, {Barrufet}, {Fujimoto}, {Heintz},
  {Nelson}, {Strait}, {Suess}, \& {Gibson}}]{Gimenez2023}
{Gim{\'e}nez-Arteaga}, C., {Oesch}, P.~A., {Brammer}, G.~B., {et~al.} 2023,
  \apj, 948, 126

\bibitem[{{Guo} {et~al.}(2015){Guo}, {Ferguson}, {Bell}, {Koo}, {Conselice},
  {Giavalisco}, {Kassin}, {Lu}, {Lucas}, {Mandelker}, {McIntosh}, {Primack},
  {Ravindranath}, {Barro}, {Ceverino}, {Dekel}, {Faber}, {Fang}, {Koekemoer},
  {Noeske}, {Rafelski}, \& {Straughn}}]{Guo2015}
{Guo}, Y., {Ferguson}, H.~C., {Bell}, E.~F., {et~al.} 2015, \apj, 800, 39

\bibitem[{{Heckman} {et~al.}(2015){Heckman}, {Alexandroff}, {Borthakur},
  {Overzier}, \& {Leitherer}}]{Heckman2015}
{Heckman}, T.~M., {Alexandroff}, R.~M., {Borthakur}, S., {Overzier}, R., \&
  {Leitherer}, C. 2015, \apj, 809, 147

\bibitem[{{Heintz} {et~al.}(2022){Heintz}, {Oesch}, {Aravena}, {Bouwens},
  {Dayal}, {Ferrara}, {Fudamoto}, {Graziani}, {Inami}, {Sommovigo}, {Smit},
  {Stefanon}, {Topping}, {Pallottini}, \& {van der Werf}}]{Heintz2022}
{Heintz}, K.~E., {Oesch}, P.~A., {Aravena}, M., {et~al.} 2022, \apjl, 934, L27

\bibitem[{{Herrera-Camus} {et~al.}(2022){Herrera-Camus}, {F{\"o}rster
  Schreiber}, {Price}, {{\"U}bler}, {Bolatto}, {Davies}, {Fisher}, {Genzel},
  {Lutz}, {Naab}, {Nestor}, {Shimizu}, {Sternberg}, {Tacconi}, \&
  {Tadaki}}]{Herrera2022}
{Herrera-Camus}, R., {F{\"o}rster Schreiber}, N.~M., {Price}, S.~H., {et~al.}
  2022, \aap, 665, L8

\bibitem[{{Hung} {et~al.}(2016){Hung}, {Hayward}, {Smith}, {Ashby}, {Lanz},
  {Mart{\'\i}nez-Galarza}, {Sanders}, \& {Zezas}}]{Hung2016}
{Hung}, C.-L., {Hayward}, C.~C., {Smith}, H.~A., {et~al.} 2016, \apj, 816, 99

\bibitem[{{Hung} {et~al.}(2015){Hung}, {Rich}, {Yuan}, {Larson}, {Casey},
  {Smith}, {Sanders}, {Kewley}, \& {Hayward}}]{Hung2015}
{Hung}, C.-L., {Rich}, J.~A., {Yuan}, T., {et~al.} 2015, \apj, 803, 62

\bibitem[{{Jakobsen} {et~al.}(2022){Jakobsen}, {Ferruit}, {Alves de Oliveira},
  {Arribas}, {Bagnasco}, {Barho}, {Beck}, {Birkmann}, {B{\"o}ker}, {Bunker},
  {Charlot}, {de Jong}, {de Marchi}, {Ehrenwinkler}, {Falcolini}, {Fels},
  {Franx}, {Franz}, {Funke}, {Giardino}, {Gnata}, {Holota}, {Honnen}, {Jensen},
  {Jentsch}, {Johnson}, {Jollet}, {Karl}, {Kling}, {K{\"o}hler}, {Kolm},
  {Kumari}, {Lander}, {Lemke}, {L{\'o}pez-Caniego}, {L{\"u}tzgendorf},
  {Maiolino}, {Manjavacas}, {Marston}, {Maschmann}, {Maurer}, {Messerschmidt},
  {Moseley}, {Mosner}, {Mott}, {Muzerolle}, {Pirzkal}, {Pittet}, {Plitzke},
  {Posselt}, {Rapp}, {Rauscher}, {Rawle}, {Rix}, {R{\"o}del}, {Rumler},
  {Sabbi}, {Salvignol}, {Schmid}, {Sirianni}, {Smith}, {Strada}, {te Plate},
  {Valenti}, {Wettemann}, {Wiehe}, {Wiesmayer}, {Willott}, {Wright}, {Zeidler},
  \& {Zincke}}]{Jakobsen2022}
{Jakobsen}, P., {Ferruit}, P., {Alves de Oliveira}, C., {et~al.} 2022, \aap,
  661, A80

\bibitem[{{Johnson}(2019)}]{Johnson2019}
{Johnson}, B.~D. 2019, {SEDPY: Modules for storing and operating on
  astronomical source spectral energy distribution}, Astrophysics Source Code
  Library, record ascl:1905.026

\bibitem[{{Jones} {et~al.}(2021){Jones}, {Vergani}, {Romano}, {Ginolfi},
  {Fudamoto}, {B{\'e}thermin}, {Fujimoto}, {Lemaux}, {Morselli}, {Capak},
  {Cassata}, {Faisst}, {Le F{\`e}vre}, {Schaerer}, {Silverman}, {Yan},
  {Boquien}, {Cimatti}, {Dessauges-Zavadsky}, {Ibar}, {Maiolino}, {Rizzo},
  {Talia}, \& {Zamorani}}]{Jones2021}
{Jones}, G.~C., {Vergani}, D., {Romano}, M., {et~al.} 2021, \mnras, 507, 3540

\bibitem[{{Kennicutt}(1998)}]{Kennicutt1998}
{Kennicutt}, Robert~C., J. 1998, \araa, 36, 189

\bibitem[{{Kennicutt} \& {Evans}(2012)}]{Kennicutt2012}
{Kennicutt}, R.~C. \& {Evans}, N.~J. 2012, \araa, 50, 531

\bibitem[{{Kohandel} {et~al.}(2019){Kohandel}, {Pallottini}, {Ferrara},
  {Zanella}, {Behrens}, {Carniani}, {Gallerani}, \& {Vallini}}]{Kohandel2019}
{Kohandel}, M., {Pallottini}, A., {Ferrara}, A., {et~al.} 2019, \mnras, 487,
  3007

\bibitem[{{Krajnovi{\'c}} {et~al.}(2006){Krajnovi{\'c}}, {Cappellari}, {de
  Zeeuw}, \& {Copin}}]{Krajnovic2006}
{Krajnovi{\'c}}, D., {Cappellari}, M., {de Zeeuw}, P.~T., \& {Copin}, Y. 2006,
  \mnras, 366, 787

\bibitem[{{Krumholz} {et~al.}(2018){Krumholz}, {Burkhart}, {Forbes}, \&
  {Crocker}}]{Krumholz2018}
{Krumholz}, M.~R., {Burkhart}, B., {Forbes}, J.~C., \& {Crocker}, R.~M. 2018,
  \mnras, 477, 2716

\bibitem[{{Lelli} {et~al.}(2021){Lelli}, {Di Teodoro}, {Fraternali}, {Man},
  {Zhang}, {De Breuck}, {Davis}, \& {Maiolino}}]{Lelli2021}
{Lelli}, F., {Di Teodoro}, E.~M., {Fraternali}, F., {et~al.} 2021, Science,
  371, 713

\bibitem[{{L{\"u}tzgendorf} {et~al.}(2022){L{\"u}tzgendorf}, {Giardino}, {Alves
  de Oliveira}, {Zeidler}, {Ferruit}, {Jakobsen}, {Kumari}, {Rawle},
  {Birkmann}, {B{\"o}ker}, {Proffitt}, {Sirianni}, {Te Plate}, \&
  {Sohn}}]{Luetzgendorf2022}
{L{\"u}tzgendorf}, N., {Giardino}, G., {Alves de Oliveira}, C., {et~al.} 2022,
  in Society of Photo-Optical Instrumentation Engineers (SPIE) Conference
  Series, Vol. 12180, Space Telescopes and Instrumentation 2022: Optical,
  Infrared, and Millimeter Wave, ed. L.~E. {Coyle}, S.~{Matsuura}, \& M.~D.
  {Perrin}, 121800Y

\bibitem[{{Mandelker} {et~al.}(2014){Mandelker}, {Dekel}, {Ceverino}, {Tweed},
  {Moody}, \& {Primack}}]{Mandelker2014}
{Mandelker}, N., {Dekel}, A., {Ceverino}, D., {et~al.} 2014, \mnras, 443, 3675

\bibitem[{{Maraston} {et~al.}(2010){Maraston}, {Pforr}, {Renzini}, {Daddi},
  {Dickinson}, {Cimatti}, \& {Tonini}}]{Maraston2010}
{Maraston}, C., {Pforr}, J., {Renzini}, A., {et~al.} 2010, \mnras, 407, 830

\bibitem[{{Maseda} {et~al.}(2013){Maseda}, {van der Wel}, {da Cunha}, {Rix},
  {Pacifici}, {Momcheva}, {Brammer}, {Franx}, {van Dokkum}, {Bell},
  {Fumagalli}, {Grogin}, {Kocevski}, {Koekemoer}, {Lundgren}, {Marchesini},
  {Nelson}, {Patel}, {Skelton}, {Straughn}, {Trump}, {Weiner}, {Whitaker}, \&
  {Wuyts}}]{Maseda2013}
{Maseda}, M.~V., {van der Wel}, A., {da Cunha}, E., {et~al.} 2013, \apjl, 778,
  L22

\bibitem[{{Moster} {et~al.}(2018){Moster}, {Naab}, \& {White}}]{Moster2018}
{Moster}, B.~P., {Naab}, T., \& {White}, S. D.~M. 2018, \mnras, 477, 1822

\bibitem[{{Narayanan} {et~al.}(2023){Narayanan}, {Lower}, {Torrey}, {Brammer},
  {Cui}, {Dave}, {Iyer}, {Li}, {Lovell}, {Sales}, {Stark}, {Marinacci}, \&
  {Vogelsberger}}]{Narayanan2023}
{Narayanan}, D., {Lower}, S., {Torrey}, P., {et~al.} 2023, arXiv e-prints,
  arXiv:2306.10118

\bibitem[{{Neeleman} {et~al.}(2020){Neeleman}, {Prochaska}, {Kanekar}, \&
  {Rafelski}}]{Neeleman2020}
{Neeleman}, M., {Prochaska}, J.~X., {Kanekar}, N., \& {Rafelski}, M. 2020,
  \nat, 581, 269

\bibitem[{{Newman} {et~al.}(2018){Newman}, {Belli}, {Ellis}, \&
  {Patel}}]{Newman2018}
{Newman}, A.~B., {Belli}, S., {Ellis}, R.~S., \& {Patel}, S.~G. 2018, \apj,
  862, 126

\bibitem[{{Nidever} {et~al.}(2023){Nidever}, {Gilbert}, {Tollerud}, {Siders},
  {Escala}, {Allende Prieto}, {Smith}, {Cunha}, {Debattista}, {Ting}, \&
  {Kirby}}]{Nidever2023}
{Nidever}, D.~L., {Gilbert}, K., {Tollerud}, E., {et~al.} 2023, arXiv e-prints,
  arXiv:2306.04688

\bibitem[{{Oesch} {et~al.}(2023){Oesch}, {Brammer}, {Naidu}, {Bouwens},
  {Chisholm}, {Illingworth}, {Matthee}, {Nelson}, {Qin}, {Reddy}, {Shapley},
  {Shivaei}, {van Dokkum}, {Weibel}, {Whitaker}, {Wuyts}, {Covelo-Paz},
  {Endsley}, {Fudamoto}, {Giovinazzo}, {Herard-Demanche}, {Kerutt},
  {Kramarenko}, {Labbe}, {Leonova}, {Lin}, {Magee}, {Marchesini}, {Maseda},
  {Mason}, {Matharu}, {Meyer}, {Neufeld}, {Prieto Lyon}, {Schaerer}, {Sharma},
  {Shuntov}, {Smit}, {Stefanon}, {Wyithe}, \& {Xiao}}]{FRESCO2023}
{Oesch}, P.~A., {Brammer}, G., {Naidu}, R.~P., {et~al.} 2023, arXiv e-prints,
  arXiv:2304.02026

\bibitem[{{O'Leary} {et~al.}(2021){O'Leary}, {Moster}, {Naab}, \&
  {Somerville}}]{OLeary2021}
{O'Leary}, J.~A., {Moster}, B.~P., {Naab}, T., \& {Somerville}, R.~S. 2021,
  \mnras, 501, 3215

\bibitem[{{Oser} {et~al.}(2010){Oser}, {Ostriker}, {Naab}, {Johansson}, \&
  {Burkert}}]{Oser2010}
{Oser}, L., {Ostriker}, J.~P., {Naab}, T., {Johansson}, P.~H., \& {Burkert}, A.
  2010, \apj, 725, 2312

\bibitem[{{Ostriker} \& {Shetty}(2011)}]{Ostriker2011}
{Ostriker}, E.~C. \& {Shetty}, R. 2011, \apj, 731, 41

\bibitem[{{Parlanti} {et~al.}(2023){Parlanti}, {Carniani}, {Pallottini},
  {Cignoni}, {Cresci}, {Kohandel}, {Mannucci}, \& {Marconi}}]{Parlanti2023}
{Parlanti}, E., {Carniani}, S., {Pallottini}, A., {et~al.} 2023, \aap, 673,
  A153

\bibitem[{{P{\'e}rez-Gonz{\'a}lez} {et~al.}(2023){P{\'e}rez-Gonz{\'a}lez},
  {Barro}, {Annunziatella}, {Costantin}, {Garc{\'\i}a-Argum{\'a}nez},
  {McGrath}, {M{\'e}rida}, {Zavala}, {Haro}, {Bagley}, {Backhaus}, {Behroozi},
  {Bell}, {Bisigello}, {Buat}, {Calabr{\`o}}, {Casey}, {Cleri}, {Coogan},
  {Cooper}, {Cooray}, {Dekel}, {Dickinson}, {Elbaz}, {Ferguson}, {Finkelstein},
  {Fontana}, {Franco}, {Gardner}, {Giavalisco}, {G{\'o}mez-Guijarro},
  {Grazian}, {Grogin}, {Guo}, {Huertas-Company}, {Jogee}, {Kartaltepe},
  {Kewley}, {Kirkpatrick}, {Kocevski}, {Koekemoer}, {Long}, {Lotz}, {Lucas},
  {Papovich}, {Pirzkal}, {Ravindranath}, {Somerville}, {Tacchella}, {Trump},
  {Wang}, {Wilkins}, {Wuyts}, {Yang}, \& {Yung}}]{Perez2023}
{P{\'e}rez-Gonz{\'a}lez}, P.~G., {Barro}, G., {Annunziatella}, M., {et~al.}
  2023, \apjl, 946, L16

\bibitem[{{Perrin} {et~al.}(2012){Perrin}, {Soummer}, {Elliott}, {Lallo}, \&
  {Sivaramakrishnan}}]{Perrin2012}
{Perrin}, M.~D., {Soummer}, R., {Elliott}, E.~M., {Lallo}, M.~D., \&
  {Sivaramakrishnan}, A. 2012, in Society of Photo-Optical Instrumentation
  Engineers (SPIE) Conference Series, Vol. 8442, Space Telescopes and
  Instrumentation 2012: Optical, Infrared, and Millimeter Wave, ed. M.~C.
  {Clampin}, G.~G. {Fazio}, H.~A. {MacEwen}, \& J.~{Oschmann}, Jacobus~M.,
  84423D

\bibitem[{{Pforr} {et~al.}(2012){Pforr}, {Maraston}, \& {Tonini}}]{Pforr2012}
{Pforr}, J., {Maraston}, C., \& {Tonini}, C. 2012, \mnras, 422, 3285

\bibitem[{{Pillepich} {et~al.}(2019){Pillepich}, {Nelson}, {Springel},
  {Pakmor}, {Torrey}, {Weinberger}, {Vogelsberger}, {Marinacci}, {Genel}, {van
  der Wel}, \& {Hernquist}}]{Pillepich2019}
{Pillepich}, A., {Nelson}, D., {Springel}, V., {et~al.} 2019, \mnras, 490, 3196

\bibitem[{{Piqu{\'e}ras} {et~al.}(2008){Piqu{\'e}ras}, {Legay}, {Legros},
  {Ferruit}, {Pecontal}, {Gnata}, \& {Mosner}}]{Piqueras2008}
{Piqu{\'e}ras}, L., {Legay}, P.~J., {Legros}, E., {et~al.} 2008, in Society of
  Photo-Optical Instrumentation Engineers (SPIE) Conference Series, Vol. 7017,
  Modeling, Systems Engineering, and Project Management for Astronomy III, ed.
  G.~Z. {Angeli} \& M.~J. {Cullum}, 70170Z

\bibitem[{{Piqu{\'e}ras} {et~al.}(2010){Piqu{\'e}ras}, {Legros}, {Pons},
  {Legay}, {Ferruit}, {Dorner}, {P{\'e}contal}, {Gnata}, \&
  {Mosner}}]{Piqueras2010}
{Piqu{\'e}ras}, L., {Legros}, E., {Pons}, A., {et~al.} 2010, in Society of
  Photo-Optical Instrumentation Engineers (SPIE) Conference Series, Vol. 7738,
  Modeling, Systems Engineering, and Project Management for Astronomy IV, ed.
  G.~Z. {Angeli} \& P.~{Dierickx}, 773812

\bibitem[{{Pope} {et~al.}(2023){Pope}, {McKinney}, {Kamieneski}, {Battisti},
  {Aretxaga}, {Brammer}, {Diego}, {Hughes}, {Keller}, {Marchesini}, {Mizener},
  {Monta{\~n}a}, {Murphy}, {Whitaker}, {Wilson}, \& {Yun}}]{Pope2023}
{Pope}, A., {McKinney}, J., {Kamieneski}, P., {et~al.} 2023, \apjl, 951, L46

\bibitem[{{Price} {et~al.}(2020){Price}, {Kriek}, {Barro}, {Shapley}, {Reddy},
  {Freeman}, {Coil}, {Shivaei}, {Azadi}, {de Groot}, {Siana}, {Mobasher},
  {Sanders}, {Leung}, {Fetherolf}, {Zick}, {{\"U}bler}, \& {F{\"o}rster
  Schreiber}}]{Price2020}
{Price}, S.~H., {Kriek}, M., {Barro}, G., {et~al.} 2020, \apj, 894, 91

\bibitem[{{Price} {et~al.}(2016){Price}, {Kriek}, {Shapley}, {Reddy},
  {Freeman}, {Coil}, {de Groot}, {Shivaei}, {Siana}, {Azadi}, {Barro},
  {Mobasher}, {Sanders}, \& {Zick}}]{Price2016}
{Price}, S.~H., {Kriek}, M., {Shapley}, A.~E., {et~al.} 2016, \apj, 819, 80

\bibitem[{{Price} {et~al.}(2022){Price}, {{\"U}bler}, {F{\"o}rster Schreiber},
  {de Zeeuw}, {Burkert}, {Genzel}, {Tacconi}, {Davies}, \& {Price}}]{Price2022}
{Price}, S.~H., {{\"U}bler}, H., {F{\"o}rster Schreiber}, N.~M., {et~al.} 2022,
  \aap, 665, A159

\bibitem[{{Rieke} {et~al.}(2023{\natexlab{a}}){Rieke}, {Robertson},
  {Tacchella}, {Hainline}, {Johnson}, {Hausen}, {Ji}, {Willmer}, \& {the JADES
  Collaboration}}]{Rieke2023}
{Rieke}, M., {Robertson}, B., {Tacchella}, S., {et~al.} 2023{\natexlab{a}},
  arXiv e-prints, arXiv:2306.02466

\bibitem[{{Rieke} {et~al.}(2023{\natexlab{b}}){Rieke}, {Kelly}, {Misselt},
  {Stansberry}, {Boyer}, {Beatty}, {Egami}, {Florian}, {Greene}, {Hainline},
  {Leisenring}, {Roellig}, {Schlawin}, {Sun}, {Tinnin}, {Williams}, {Willmer},
  {Wilson}, {Clark}, {Rohrbach}, {Brooks}, {Canipe}, {Correnti}, {DiFelice},
  {Gennaro}, {Girard}, {Hartig}, {Hilbert}, {Koekemoer}, {Nikolov}, {Pirzkal},
  {Rest}, {Robberto}, {Sunnquist}, {Telfer}, {Wu}, {Ferry}, {Lewis}, {Baum},
  {Beichman}, {Doyon}, {Dressler}, {Eisenstein}, {Ferrarese}, {Hodapp},
  {Horner}, {Jaffe}, {Johnstone}, {Krist}, {Martin}, {McCarthy}, {Meyer},
  {Rieke}, {Trauger}, \& {Young}}]{NIRCamInst}
{Rieke}, M.~J., {Kelly}, D.~M., {Misselt}, K., {et~al.} 2023{\natexlab{b}},
  \pasp, 135, 028001

\bibitem[{{Rigby} {et~al.}(2023){Rigby}, {Perrin}, {McElwain}, {Kimble},
  {Friedman}, {Lallo}, {Doyon}, {Feinberg}, {Ferruit}, {Glasse}, {Rieke},
  {Rieke}, {Wright}, {Willott}, {Colon}, {Milam}, {Neff}, {Stark}, {Valenti},
  {Abell}, {Abney}, {Abul-Huda}, {Acton}, {Adams}, {Adler}, {Aguilar}, {Ahmed},
  {Albert}, {Alberts}, {Aldridge}, {Allen}, {Altenburg},
  {{\'A}lvarez-M{\'a}rquez}, {Alves de Oliveira}, {Andersen}, {Anderson},
  {Anderson}, {Argyriou}, \& {Armstrong}}]{Rigby2023}
{Rigby}, J., {Perrin}, M., {McElwain}, M., {et~al.} 2023, \pasp, 135, 048001

\bibitem[{{Rizzo} {et~al.}(2023){Rizzo}, {Roman-Oliveira}, {Fraternali},
  {Frickmann}, {Valentino}, {Brammer}, {Zanella}, {Kokorev}, {Popping},
  {Whitaker}, {Kohandel}, {Magdis}, {Di Mascolo}, {Ikeda}, {Jin}, \&
  {Toft}}]{Rizzo2023}
{Rizzo}, F., {Roman-Oliveira}, F., {Fraternali}, F., {et~al.} 2023, arXiv
  e-prints, arXiv:2303.16227

\bibitem[{{Rizzo} {et~al.}(2021){Rizzo}, {Vegetti}, {Fraternali}, {Stacey}, \&
  {Powell}}]{Rizzo2021}
{Rizzo}, F., {Vegetti}, S., {Fraternali}, F., {Stacey}, H.~R., \& {Powell}, D.
  2021, \mnras, 507, 3952

\bibitem[{{Rizzo} {et~al.}(2020){Rizzo}, {Vegetti}, {Powell}, {Fraternali},
  {McKean}, {Stacey}, \& {White}}]{Rizzo2020}
{Rizzo}, F., {Vegetti}, S., {Powell}, D., {et~al.} 2020, \nat, 584, 201

\bibitem[{{Robertson} {et~al.}(2006){Robertson}, {Bullock}, {Cox}, {Di Matteo},
  {Hernquist}, {Springel}, \& {Yoshida}}]{Robertson2006}
{Robertson}, B., {Bullock}, J.~S., {Cox}, T.~J., {et~al.} 2006, \apj, 645, 986

\bibitem[{{Rodrigues} {et~al.}(2017){Rodrigues}, {Hammer}, {Flores}, {Puech},
  \& {Athanassoula}}]{Rodrigues2017}
{Rodrigues}, M., {Hammer}, F., {Flores}, H., {Puech}, M., \& {Athanassoula}, E.
  2017, \mnras, 465, 1157

\bibitem[{{Rodriguez-Gomez} {et~al.}(2015){Rodriguez-Gomez}, {Genel},
  {Vogelsberger}, {Sijacki}, {Pillepich}, {Sales}, {Torrey}, {Snyder},
  {Nelson}, {Springel}, {Ma}, \& {Hernquist}}]{Rodriguez2015}
{Rodriguez-Gomez}, V., {Genel}, S., {Vogelsberger}, M., {et~al.} 2015, \mnras,
  449, 49

\bibitem[{{Sattari} {et~al.}(2023){Sattari}, {Mobasher}, {Chartab}, {Kelson},
  {Teplitz}, {Rafelski}, {Grogin}, {Koekemoer}, {Wang}, {Windhorst}, {Alavi},
  {Prichard}, {Sunnquist}, {Gardner}, {Gawiser}, {Hathi}, {Hayes}, {Ji},
  {Mehta}, {Robertson}, {Scarlata}, {Yung}, {Conselice}, {Dai}, {Guo}, {Lucas},
  {Martin}, \& {Ravindranath}}]{Sattari2023}
{Sattari}, Z., {Mobasher}, B., {Chartab}, N., {et~al.} 2023, \apj, 951, 147

\bibitem[{{S\'ersic}(1968)}]{Sersic1968}
{S\'ersic}, J.~L. 1968, {Atlas de Galaxias Australes}

\bibitem[{{Shapiro} {et~al.}(2008){Shapiro}, {Genzel}, {F{\"o}rster Schreiber},
  {Tacconi}, {Bouch{\'e}}, {Cresci}, {Davies}, {Eisenhauer}, {Johansson},
  {Krajnovi{\'c}}, {Lutz}, {Naab}, {Arimoto}, {Arribas}, {Cimatti}, {Colina},
  {Daddi}, {Daigle}, {Erb}, {Hernandez}, {Kong}, {Mignoli}, {Onodera},
  {Renzini}, {Shapley}, \& {Steidel}}]{Shapiro2008}
{Shapiro}, K.~L., {Genzel}, R., {F{\"o}rster Schreiber}, N.~M., {et~al.} 2008,
  \apj, 682, 231

\bibitem[{{Shetty} \& {Ostriker}(2012)}]{Shetty2012}
{Shetty}, R. \& {Ostriker}, E.~C. 2012, \apj, 754, 2

\bibitem[{{Simons} {et~al.}(2019){Simons}, {Kassin}, {Snyder}, {Primack},
  {Ceverino}, {Dekel}, {Hayward}, {Mandelker}, {Mantha}, {Pacifici}, {de la
  Vega}, \& {Wang}}]{Simons2019}
{Simons}, R.~C., {Kassin}, S.~A., {Snyder}, G.~F., {et~al.} 2019, \apj, 874, 59

\bibitem[{{Simons} {et~al.}(2017){Simons}, {Kassin}, {Weiner}, {Faber},
  {Trump}, {Heckman}, {Koo}, {Pacifici}, {Primack}, {Snyder}, \& {de la
  Vega}}]{Simons2017}
{Simons}, R.~C., {Kassin}, S.~A., {Weiner}, B.~J., {et~al.} 2017, \apj, 843, 46

\bibitem[{{Sorba} \& {Sawicki}(2018)}]{Sorba2018}
{Sorba}, R. \& {Sawicki}, M. 2018, \mnras, 476, 1532

\bibitem[{{Stott} {et~al.}(2016){Stott}, {Swinbank}, {Johnson}, {Tiley},
  {Magdis}, {Bower}, {Bunker}, {Bureau}, {Harrison}, {Jarvis}, {Sharples},
  {Smail}, {Sobral}, {Best}, \& {Cirasuolo}}]{Stott2016}
{Stott}, J.~P., {Swinbank}, A.~M., {Johnson}, H.~L., {et~al.} 2016, \mnras,
  457, 1888

\bibitem[{{Suess} {et~al.}(2023){Suess}, {Williams}, {Robertson}, {Ji},
  {Johnson}, {Nelson}, {Alberts}, {Hainline}, {DEugenio}, {Ubler}, {Rieke},
  {Rieke}, {Bunker}, {Carniani}, {Charlot}, {Eisenstein}, {Maiolino}, {Stark},
  {Tacchella}, \& {Willott}}]{Suess2023}
{Suess}, K.~A., {Williams}, C.~C., {Robertson}, B., {et~al.} 2023, arXiv
  e-prints, arXiv:2307.14209

\bibitem[{{Tacchella} {et~al.}(2023){Tacchella}, {Johnson}, {Robertson},
  {Carniani}, {D'Eugenio}, {Kumari}, {Maiolino}, {Nelson}, {Suess},
  {{\"U}bler}, {Williams}, {Adebusola}, {Alberts}, {Arribas}, {Bhatawdekar},
  {Bonaventura}, {Bowler}, {Bunker}, {Cameron}, {Curti}, {Egami}, {Eisenstein},
  {Frye}, {Hainline}, {Helton}, {Ji}, {Looser}, {Lyu}, {Perna}, {Rawle},
  {Rieke}, {Rieke}, {Saxena}, {Sandles}, {Shivaei}, {Simmonds}, {Sun},
  {Willmer}, {Willott}, \& {Witstok}}]{Tacchella2023}
{Tacchella}, S., {Johnson}, B.~D., {Robertson}, B.~E., {et~al.} 2023, \mnras,
  522, 6236

\bibitem[{{Tacconi} {et~al.}(2018){Tacconi}, {Genzel}, {Saintonge}, {Combes},
  {Garc{\'\i}a-Burillo}, {Neri}, {Bolatto}, {Contini}, {F{\"o}rster Schreiber},
  {Lilly}, {Lutz}, {Wuyts}, {Accurso}, {Boissier}, {Boone}, {Bouch{\'e}},
  {Bournaud}, {Burkert}, {Carollo}, {Cooper}, {Cox}, {Feruglio}, {Freundlich},
  {Herrera-Camus}, {Juneau}, {Lippa}, {Naab}, {Renzini}, {Salome}, {Sternberg},
  {Tadaki}, {{\"U}bler}, {Walter}, {Weiner}, \& {Weiss}}]{Tacconi2018}
{Tacconi}, L.~J., {Genzel}, R., {Saintonge}, A., {et~al.} 2018, \apj, 853, 179

\bibitem[{{Tacconi} {et~al.}(2020){Tacconi}, {Genzel}, \&
  {Sternberg}}]{Tacconi2020}
{Tacconi}, L.~J., {Genzel}, R., \& {Sternberg}, A. 2020, \araa, 58, 157

\bibitem[{{Topping} {et~al.}(2022){Topping}, {Stark}, {Endsley}, {Bouwens},
  {Schouws}, {Smit}, {Stefanon}, {Inami}, {Bowler}, {Oesch}, {Gonzalez},
  {Dayal}, {da Cunha}, {Algera}, {van der Werf}, {Pallottini}, {Barrufet},
  {Schneider}, {De Looze}, {Sommovigo}, {Whitler}, {Graziani}, {Fudamoto}, \&
  {Ferrara}}]{Topping2022}
{Topping}, M.~W., {Stark}, D.~P., {Endsley}, R., {et~al.} 2022, \mnras, 516,
  975

\bibitem[{{Turner} {et~al.}(2017){Turner}, {Cirasuolo}, {Harrison}, {McLure},
  {Dunlop}, {Swinbank}, {Johnson}, {Sobral}, {Matthee}, \&
  {Sharples}}]{Turner2017}
{Turner}, O.~J., {Cirasuolo}, M., {Harrison}, C.~M., {et~al.} 2017, \mnras,
  471, 1280

\bibitem[{{{\"U}bler} {et~al.}(2019){{\"U}bler}, {Genzel}, {Wisnioski},
  {F{\"o}rster Schreiber}, {Shimizu}, {Price}, {Tacconi}, {Belli}, {Wilman},
  {Fossati}, {Mendel}, {Davies}, {Beifiori}, {Bender}, {Brammer}, {Burkert},
  {Chan}, {Davies}, {Fabricius}, {Galametz}, {Herrera-Camus}, {Lang}, {Lutz},
  {Momcheva}, {Naab}, {Nelson}, {Saglia}, {Tadaki}, {van Dokkum}, \&
  {Wuyts}}]{Uebler2019}
{{\"U}bler}, H., {Genzel}, R., {Wisnioski}, E., {et~al.} 2019, \apj, 880, 48

\bibitem[{{van de Sande} {et~al.}(2018){van de Sande}, {Scott},
  {Bland-Hawthorn}, {Brough}, {Bryant}, {Colless}, {Cortese}, {Croom},
  {d'Eugenio}, {Foster}, {Goodwin}, {Konstantopoulos}, {Lawrence}, {McDermid},
  {Medling}, {Owers}, {Richards}, \& {Sharp}}]{vdSande2018}
{van de Sande}, J., {Scott}, N., {Bland-Hawthorn}, J., {et~al.} 2018, Nature
  Astronomy, 2, 483

\bibitem[{{van der Wel} {et~al.}(2012){van der Wel}, {Bell}, {H{\"a}ussler},
  {McGrath}, {Chang}, {Guo}, {McIntosh}, {Rix}, {Barden}, {Cheung}, {Faber},
  {Ferguson}, {Galametz}, {Grogin}, {Hartley}, {Kartaltepe}, {Kocevski},
  {Koekemoer}, {Lotz}, {Mozena}, {Peth}, \& {Peng}}]{vdWel2012}
{van der Wel}, A., {Bell}, E.~F., {H{\"a}ussler}, B., {et~al.} 2012, \apjs,
  203, 24

\bibitem[{{van der Wel} {et~al.}(2014){van der Wel}, {Chang}, {Bell}, {Holden},
  {Ferguson}, {Giavalisco}, {Rix}, {Skelton}, {Whitaker}, {Momcheva},
  {Brammer}, {Kassin}, {Martig}, {Dekel}, {Ceverino}, {Koo}, {Mozena}, {van
  Dokkum}, {Franx}, {Faber}, \& {Primack}}]{vdWel2014}
{van der Wel}, A., {Chang}, Y.-Y., {Bell}, E.~F., {et~al.} 2014, \apjl, 792, L6

\bibitem[{{van Dokkum} {et~al.}(2015){van Dokkum}, {Nelson}, {Franx}, {Oesch},
  {Momcheva}, {Brammer}, {F{\"o}rster Schreiber}, {Skelton}, {Whitaker}, {van
  der Wel}, {Bezanson}, {Fumagalli}, {Illingworth}, {Kriek}, {Leja}, \&
  {Wuyts}}]{vDokkum2015}
{van Dokkum}, P.~G., {Nelson}, E.~J., {Franx}, M., {et~al.} 2015, \apj, 813, 23

\bibitem[{{van Houdt} {et~al.}(2021){van Houdt}, {van der Wel}, {Bezanson},
  {Franx}, {d'Eugenio}, {Barisic}, {Bell}, {Gallazzi}, {de Graaff}, {Maseda},
  {Pacifici}, {van de Sande}, {Sobral}, {Straatman}, \& {Wu}}]{vanHoudt2021}
{van Houdt}, J., {van der Wel}, A., {Bezanson}, R., {et~al.} 2021, \apj, 923,
  11

\bibitem[{{Whitaker} {et~al.}(2014){Whitaker}, {Franx}, {Leja}, {van Dokkum},
  {Henry}, {Skelton}, {Fumagalli}, {Momcheva}, {Brammer}, {Labb{\'e}},
  {Nelson}, \& {Rigby}}]{Whitaker2014}
{Whitaker}, K.~E., {Franx}, M., {Leja}, J., {et~al.} 2014, \apj, 795, 104

\bibitem[{{White} \& {Rees}(1978)}]{White1978}
{White}, S.~D.~M. \& {Rees}, M.~J. 1978, \mnras, 183, 341

\bibitem[{{Whitler} {et~al.}(2023){Whitler}, {Stark}, {Endsley}, {Leja},
  {Charlot}, \& {Chevallard}}]{Whitler2023}
{Whitler}, L., {Stark}, D.~P., {Endsley}, R., {et~al.} 2023, \mnras, 519, 5859

\bibitem[{{Wisnioski} {et~al.}(2019){Wisnioski}, {F{\"o}rster Schreiber},
  {Fossati}, {Mendel}, {Wilman}, {Genzel}, {Bender}, {Wuyts}, {Davies},
  {{\"U}bler}, {Bandara}, {Beifiori}, {Belli}, {Brammer}, {Chan}, {Davies},
  {Fabricius}, {Galametz}, {Lang}, {Lutz}, {Nelson}, {Momcheva}, {Price},
  {Rosario}, {Saglia}, {Seitz}, {Shimizu}, {Tacconi}, {Tadaki}, {van Dokkum},
  \& {Wuyts}}]{Wisnioski2019}
{Wisnioski}, E., {F{\"o}rster Schreiber}, N.~M., {Fossati}, M., {et~al.} 2019,
  \apj, 886, 124

\bibitem[{{Wisnioski} {et~al.}(2015){Wisnioski}, {F{\"o}rster Schreiber},
  {Wuyts}, {Wuyts}, {Bandara}, {Wilman}, {Genzel}, {Bender}, {Davies},
  {Fossati}, {Lang}, {Mendel}, {Beifiori}, {Brammer}, {Chan}, {Fabricius},
  {Fudamoto}, {Kulkarni}, {Kurk}, {Lutz}, {Nelson}, {Momcheva}, {Rosario},
  {Saglia}, {Seitz}, {Tacconi}, \& {van Dokkum}}]{Wisnioski2015}
{Wisnioski}, E., {F{\"o}rster Schreiber}, N.~M., {Wuyts}, S., {et~al.} 2015,
  \apj, 799, 209

\bibitem[{{Wuyts} {et~al.}(2016){Wuyts}, {F{\"o}rster Schreiber}, {Wisnioski},
  {Genzel}, {Burkert}, {Bandara}, {Beifiori}, {Belli}, {Bender}, {Brammer},
  {Chan}, {Davies}, {Fossati}, {Galametz}, {Kulkarni}, {Lang}, {Lutz},
  {Mendel}, {Momcheva}, {Naab}, {Nelson}, {Saglia}, {Seitz}, {Tacconi},
  {Tadaki}, {{\"U}bler}, {van Dokkum}, {Wilman}, \& {Wuyts}}]{Wuyts2016}
{Wuyts}, S., {F{\"o}rster Schreiber}, N.~M., {Wisnioski}, E., {et~al.} 2016,
  \apj, 831, 149

\bibitem[{{Xu} {et~al.}(2022){Xu}, {Heckman}, {Henry}, {Berg}, {Chisholm},
  {James}, {Martin}, {Stark}, {Aloisi}, {Amor{\'\i}n}, {Arellano-C{\'o}rdova},
  {Bordoloi}, {Charlot}, {Chen}, {Hayes}, {Mingozzi}, {Sugahara}, {Kewley},
  {Ouchi}, {Scarlata}, \& {Steidel}}]{Xu2022}
{Xu}, X., {Heckman}, T., {Henry}, A., {et~al.} 2022, \apj, 933, 222

\bibitem[{{Zhang} {et~al.}(2019){Zhang}, {Primack}, {Faber}, {Koo}, {Dekel},
  {Chen}, {Ceverino}, {Chang}, {Fang}, {Guo}, {Lin}, \& {Wel}}]{Zhang2019}
{Zhang}, H., {Primack}, J.~R., {Faber}, S.~M., {et~al.} 2019, \mnras, 484, 5170

\end{thebibliography}

\begin{appendix}

\section{NIRSpec resolution}\label{sec:lsf}

\begin{figure*}[!ht]
    \centering
    \includegraphics[width=0.8\linewidth]{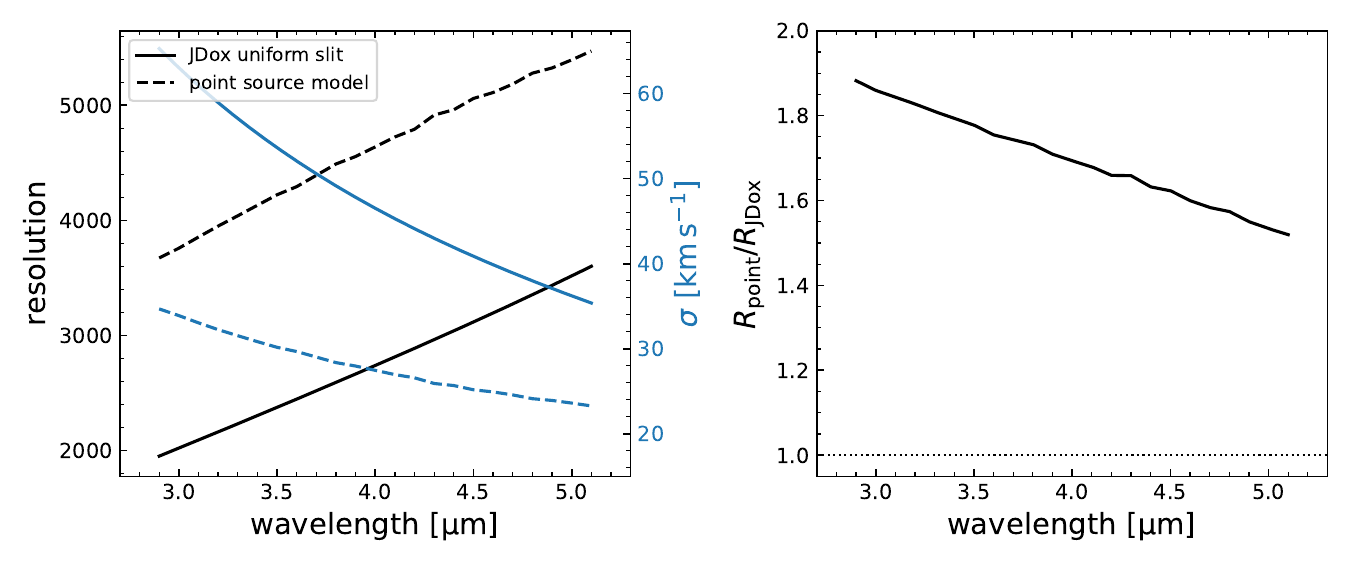}
    \caption{The resolution and line spread function (LSF) of the NIRSpec dispersers depend strongly on the light profile of the source. \textit{Left:} the resolution and corresponding velocity dispersion estimated using our modelling software for a point source in the centre of the shutter and G395H grating (dashed lines). Solid lines show the `dispersion curves' available on the JWST User Documentation webpage (JDox), which provides a (pre-launch) approximation for the LSF of a uniformly-illuminated shutter. \textit{Right:} The ratio between the resolution for a point source and LSF approximation for a uniform light distribution. At shorter wavelengths the spectral resolution is nearly a factor 2 higher for a point source. }
    \label{fig:lsf}
\end{figure*}

The instrument resolution for NIRSpec provided in the online JWST User Documentation (JDox) provides a pre-launch estimate of the 1D instrument LSF for a uniformly-illuminated slit\footnote{https://jwst-docs.stsci.edu/jwst-near-infrared-spectrograph/nirspec-instrumentation/nirspec-dispersers-and-filters}. However, as we account for the source morphology in our modelling already, the point source resolution is of relevance in our work. To provide insight into our model PSFs which are 2D images, we use our modelling software to estimate the 1D resolution as a function of wavelength for a point source.

To do so, we set up a model cube $I(x,y,\lambda)$ following the procedure described in Section~\ref{sec:method}, but for a point source morphology instead of a S\'ersic profile. We place the point source in the centre of the shutter, and use a central shutter ($s_{ij}=(185,85)$) of the fourth quadrant of the MSA. We apply a sparse wavelength sampling of $\Delta\lambda=0.2\,\micron$ across $2.7<\lambda<5.3\,\micron$ to avoid overlapping and to speed up the computation. We then propagate this model onto the detector plane and extract a 1D spectrum. By modelling the resulting emission lines at the known input wavelengths as a series of single Gaussians with variable width and amplitude, we derive the LSF for a point source. 

We show the resulting resolution and LSF for a point source for the G395H grating in Fig.~\ref{fig:lsf}. The point source resolution of NIRSpec (dashed lines) is substantially higher than the estimate for a uniformly-illuminated slit (solid lines). In the right-hand panel we show the ratio of the two as a function of wavelength: at shorter wavelength the resolution differs by nearly a factor two. At longer wavelength this discrepancy in resolution gradually decreases due to the fact that the ratio between the {spatial} PSF FWHM and the slit width increases with wavelength (i.e., the {spatial} PSF becomes broader and therefore closer to the approximation of a uniform flux distribution). We caution that this LSF cannot be applied directly to 1D rectified, extracted and combined spectra from real observations, as it does not include broadening that is usually introduced by the reduction pipeline {when resampling and combining} data from multiple exposures. In principle, our model detectors can be propagated through any reduction pipeline to estimate this additional broadening, and we investigate these effects in more detail in de Graaff et al. (in prep).

{Moreover, as noted in Section~\ref{sec:msa_description}, we emphasise that the PSF models used in our forward modelling rely partially on reference files that were created pre-launch, which likely leads to a systematic uncertainty in the FWHM. A small number of calibration programs that aim to characterise the PSF have been executed, which we discuss as follows.

For the MSA a single standard star was observed in the prism mode within the central shutter of each quadrant to obtain the spatial PSF (PID 1128). Similarly, for the fixed slit (FS; PID 1128, 1487) a standard star was observed at different intraslit positions. These data are primarily suited and used to estimate path losses. Because the PSF is highly undersampled (pixels of $0.1\arcsec$ versus a FWHM $\approx 0.03-0.16\arcsec$ if the PSF is diffraction-limited), and only one object was observed, it is extremely challenging in practice to extract a robust spatial profile from these data. \citet{Jakobsen2022} report that the spatial PSF of NIRSpec is diffraction-limited beyond $3.2\,\micron$, and therefore similar to the NIRCam PSF at long wavelengths. All emission lines considered in this paper are at $\lambda>3.4\,\micron$ and in this diffraction-limited regime.

We assess the quality of the spatial (i.e. cross-dispersion) direction of our PSF models by comparing the half-light radii inferred from our modelling to the NIRCam images and those inferred from the spectral modelling. Although we have set a prior based on the NIRCam morphological information to model the spectra, the widths of the priors are sufficiently broad. Therefore, in case of a substantial deviation between the true PSF and our model PSF, we would expect to find a systematic difference in the half-light radii between the two measurements, particularly for the objects that are compact ($<0.15\arcsec$; Table~\ref{tab:fit_results}). Yet, we find a median difference of only $\approx 10\%$, with the spectral fits indicating slightly larger sizes. This small bias may also be a physical effect, as the NIRCam images contain flux from both the stellar continuum and emission lines.

For the dispersion direction, the only available calibration data are of a distant planetary nebula (PN), which was observed in the MOS and FS modes (PID 1125; 1492). Apart from the fact that this object may not be a true point source, the undersampling again is the dominant issue in calibrating the broadening in the dispersion direction. Similar to the spatial PSF, the LSF is highly undersampled for a point source (FWHM$\approx1\,$pix). The fact that the PN spectrum shows only few ($\sim 10$) narrow emission lines across $3-5\,\micron$ and that only a single object was observed, means that constraining the spectral broadening is a major outstanding challenge. 

However, our modelling software also makes predictions for other NIRSpec dispersers. Particularly relevant is the preliminary analysis by \citet{Nidever2023}, who observed $\approx 100$ red giant stars in M31 with the G140H grating (PID 2609). These spectra show many narrow absorption line features and at high SNR, thereby enabling an empirical reconstruction of the LSF in spite of the undersampling issues. \citet{Nidever2023} find that for the G140H grating the resolution is much higher than reported on the JDox, with $R\sim4000-5000$, although the precise wavelength range for this estimate is not stated. Nevertheless, this broadly agrees with predictions from our modelling software for this grating, which suggest $R\sim4000-6000$ depending on wavelength. 
Although subject to further calibration, we conclude that our PSF models are sufficiently realistic for the purpose of this paper, and we approximate a $\approx10-20\%$ systematic uncertainty in the PSF FWHM. }

\section{{Prism spectra and SEDs}}\label{sec:beagle}

\begin{figure*}[h]
    \centering
    \includegraphics[width=\linewidth]{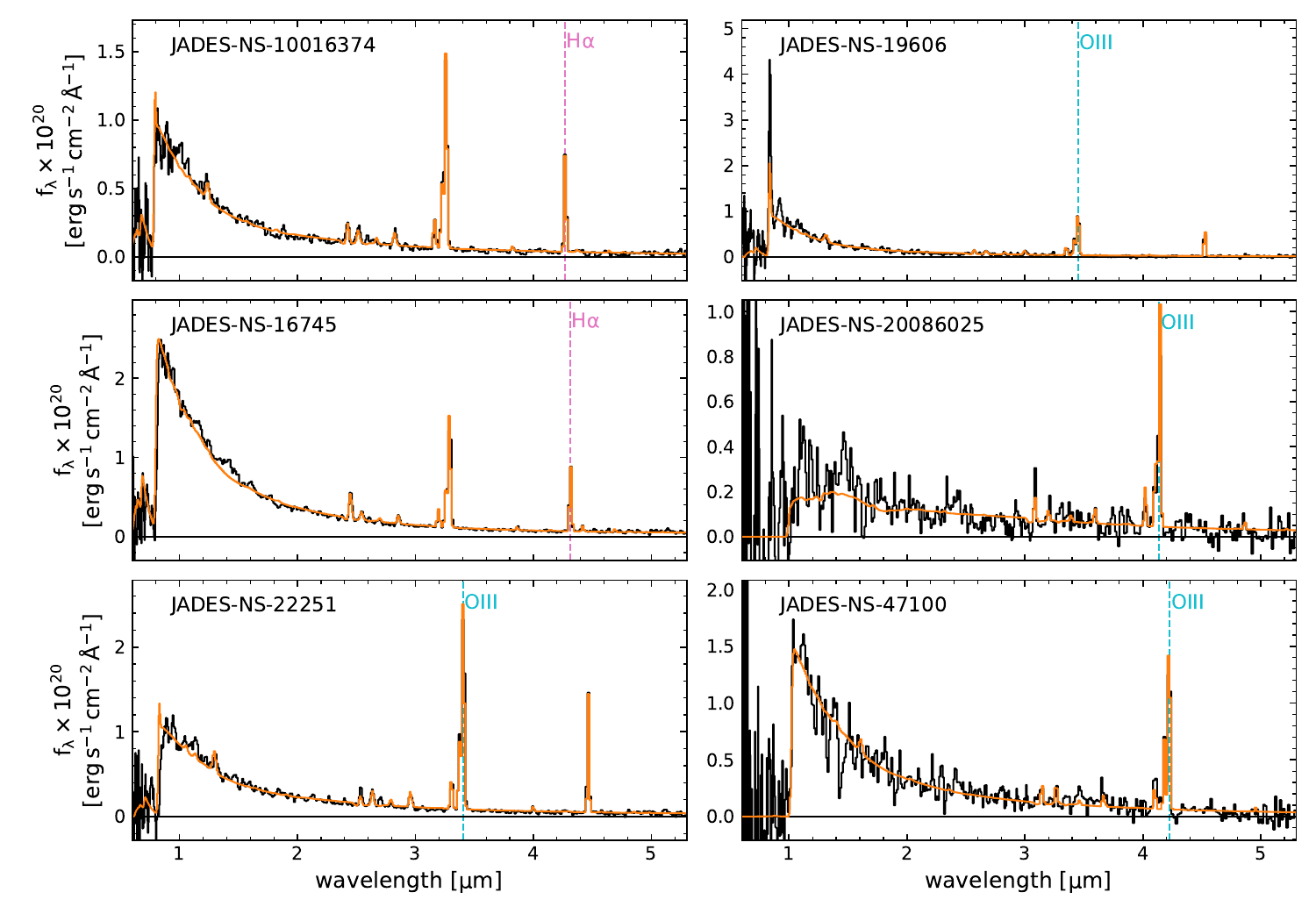}
    \caption{{Prism ($R\sim100$) spectra (black lines) and best-fit SED models estimated using BEAGLE (orange lines). The emission lines used for the dynamical modelling are highlighted by dashed lines. The SEDs of all objects are very blue and show strong emission lines, pointing to young star-forming stellar populations with low stellar masses and high specific SFRs. }}
    \label{fig:beagle_example}
\end{figure*}

As described in Section~\ref{sec:beagle}, we have performed SED modelling to the prism ($R\sim100$) spectroscopy of our sample using the BEAGLE software \citep{Chevallard2016}. {We show these spectra and best-fit (minimum $\chi^2$) models in Fig.~\ref{fig:beagle_example}. 

Generally, the spectra show very blue UV continua and strong emission lines. As a result, we infer a low stellar mass and high specific SFR (median $\rm sSFR\approx -8.0\,{yr^{-1}}$). The stellar masses and SFRs, corrected for slit losses, are presented in Table~\ref{tab:beagle}.

We also use the prism spectra to calculate the contribution of the emission lines to the NIRCam images of Fig.~\ref{fig:sample_overview}, to test our assumption that the images trace the morphology of the emission lines rather than the stellar continua. We use the \texttt{sedpy} software \citep{Johnson2019} and the observed prism spectrum to compute the total AB magnitude in the relevant NIRCam filter. Then, we construct a spectrum with only strong emission lines (H$\beta$, the [OIII] doublet, H$\alpha$), using emission line fluxes measured with pPXF as described in D'Eugenio et al. (in prep.), and compute the AB magnitude in the NIRCam filter from the emission lines only. 

We present the flux ratios in Table~\ref{tab:nircam_fluxes}. For four out of six objects, which are observed in the medium bands, the emission lines dominate the NIRCam photometry ($72$\% of the flux on average). For the other two objects observed in broad bands the emission lines do not dominate, but contribute substantially to the photometry. }

\begin{table}[h]
    \caption{BEAGLE modelling results. Values are the median of the posterior probability distributions, and uncertainties reflect the $16^{\rm th}$ and $84^{\rm th}$ percentiles.  }
    \centering
    \renewcommand{\arraystretch}{1.3}
    \begin{tabular}{l l l }
    \hline \hline
       ID &   $\log(M_*/\Msun)$ & SFR  \\
         & &  ($\Msun\,{\rm yr}^{-1}$) \\ 
        \hline
        JADES-NS-00016745 & $8.80_{-0.04}^{+0.04}$ & $12.6_{-0.8}^{+0.6}$ \\
        JADES-NS-00019606 & $7.54_{-0.04}^{+0.04}$ & $1.46_{-0.03}^{+0.03}$ \\
        JADES-NS-00022251 & $8.21_{-0.05}^{+0.05}$ & $6.2_{-0.4}^{+0.4}$ \\
        JADES-NS-00047100 & $8.53_{-0.10}^{+0.12}$ & $5.1_{-0.7}^{+0.8}$ \\        
        JADES-NS-10016374 & $7.86_{-0.07}^{+0.07}$ & $2.19_{-0.13}^{+0.12}$ \\
        JADES-NS-20086025 & $8.85_{-0.18}^{+0.18}$ & $24_{-11}^{+12}$ \\
    \hline
    \end{tabular}

    \label{tab:beagle}
\end{table}

\begin{table}[h]
    \caption{The ratio of the emission line fluxes to the total fluxes (continuum + emission lines) within the NIRCam filters used for morphological modelling.  }
    \centering
    \renewcommand{\arraystretch}{1.3}
    \begin{tabular}{l l l }
    \hline \hline
       ID &   Filter & Flux ratio  \\
        \hline
        JADES-NS-00016745 & F356W & 0.37 \\
        JADES-NS-00019606 & F335M & 0.75 \\
        JADES-NS-00022251 & F335M & 0.76 \\
        JADES-NS-00047100 & F410M & 0.69 \\        
        JADES-NS-10016374 & F444W & 0.32 \\
        JADES-NS-20086025 & F410M & 0.69 \\
    \hline
    \end{tabular}

    \label{tab:nircam_fluxes}
\end{table}

\end{appendix}

\end{document}